# Elastic Cerenkov effects in anisotropic soft materials-I: Theoretical analysis, simulations and inverse method

Guoyang Li[a], Yang Zheng[a], Yanlin Liu[a], Michel Destrade[b], Yanping Cao[a]

[a]*Institute of Biomechanics and Medical Engineering, AML, Department of Engineering Mechanics, Tsinghua University, Beijing 100084, PR China*

[b]*School of Mathematics, Statistics and Applied Mathematics, National University of Ireland Galway, Galway, Ireland*

**Abstract**

A body force concentrated at a point and moving at high speed can induce shear-wave Mach cones in dusty-plasma crystals or soft materials. These cones have been observed in experiments and the phenomenon has subsequently be named the Elastic Cerenkov Effect (ECE). The ECE in soft materials forms the basis of the supersonic shear imaging technique, an ultrasound-based dynamic elastography method applied to clinical environments in recent years. Previous studies on the ECE in soft materials focused on isotropic material models whilst here, we investigate its existence and key features in anisotropic soft media, with a view to apply the results to the non-invasive, non-destructive characterization of biological soft tissues. We focus on incompressible transversely isotropic soft materials, using both theoretical analysis and finite element simulations. We also study theoretically the characteristics of the shear wave induced in a deformed hyperelastic anisotropic soft material by a source moving with supersonic speed. Based on our theoretical analysis and numerical simulations, we propose an inverse approach to infer the linear and hyperelastic parameters of anisotropic soft materials. Finally we validate the inverse method with numerical experiments. In Part II of the paper, experiments are conducted on phantom materials to further verify the method proposed here; moreover, both *ex vivo* and *in vivo* experiments are carried out the demonstrate the usefulness of the inverse method for instance in clinical use.

# 1 Introduction

The systematic theoretical treatment on the waves generated in an infinite elastic solid by a variable body force can be dated back to the classic work of Eeason et al. (1955). In the introduction of that paper, the authors make the following remark on their theoretical solutions, which seemed reasonable at that time: "*In many cases, too, it is difficult to see how the solutions as they stand can be applied to a practical engineering problem. A body force concentrated at a point and moving with uniform velocity through an infinite solid is not easy to envisage physically.*" In fact, about 50 years after the publication of that seminal paper, experiments on wave motion in a dusty-plasma crystal clearly demonstrated shear-wave Mach cones generated by a laser radiation force moving with a high speed (Melzer et al, 2000; Nosenko et al., 2002). Then a French group (Bercoff et al., 2004a; 2004b) proposed a technique to generate a concentrated body force moving with supersonic speed through soft media in order to generate shear-wave Mach cones. Their key idea was to use ultrasonic focused beams to generate remotely a mechanical vibration source inside the soft medium. The vibration source can be moved at supersonic speed along a given direction and the resulting shear waves will interfere constructively along a Mach cone according to the theory developed by Eason et al. (1955), creating two intense plane shear waves (Fig. 1d). This phenomenon is analogous to the "sonic boom" created by a supersonic aircraft (Fig. 1a), to the Kelvin ship-wave generated by a high speed yacht (Fig. 1b) and to the Cerenkov effects induced by a high-energy charged particle passing through a transparent medium at a speed greater than the speed of light in that medium (Fig. 1c). Hence Bercoff et al. (2004a; 2004b) named their phenomenon the Elastic Cerenkov Effect (ECE, Fig. 1d), which now forms the theoretical basis of the supersonic shear imaging (SSI) technique.

The SSI technique (Bercoff et al., 2004a) represents one of the dynamic elastography methods (Sarvazyan et al., 1995, 1998) which have received considerable attention since the 1990s. Dynamic elastography methods using shear waves traveling in human soft tissues to deduce their *in vivo* elastic properties have found widespread clinical applications, including monitoring the development of liver fibrosis and the detection of malignant tumors.

Liver fibrosis is a common pathway for a multitude of liver injuries. A precise estimation of the degree of liver fibrosis is of great importance for evaluation of prognosis, surveillance, and treatment decisions in patients with chronic liver disease (Bavu et al., 2011; Ferraioli et al., 2012; Bota et al., 2013; Cassinotto et al., 2013; Paparo et al., 2014). At present, liver biopsy still serves as the reference standard for the assessment of liver fibrosis. However, it is an invasive method, associated with patient discomfort, and its accuracy is limited by intra- and inter-observer variability and sampling errors. Therefore, the development of a non-invasive method such as elastography to inspect the occurrence and development of liver fibrosis is in high demand.

Tumors are frequently detected through physical palpation as hard masses located within softer surrounding tissue. Elastography methods may serve as a "virtual finger" to quantitatively detect the hardness of the tumor, a measurement which has the potential to differentiate malignant tumors from benign ones (Bercoff et al, 2004a; Tanter et al. 2008; Athanasiou et al., 2010; Chammings et al., 2013).

The SSI technique distinguishes itself from other dynamic elastography methods by relying on the ECE induced in an isotropic soft media. Then it relies on ultrafast imaging techniques to visualize and measure the speed of the resulting shear wave in less than 20 ms (Bercoff et al., 2004a; 2004b). In principle this measurement enables us to map the elasticity of the tissue, once we relate wave speed to stiffness. This novel technique demonstrates great potential in the diagnosis of some diseases; however, some fundamental issues remain in the clinical use of the SSI technique that deserve careful investigations. This paper is concerned with the following issues.

**First**, it is well known that most soft biological tissues are anisotropic materials; this category includes the cardiovascular system, skin, kidneys and muscles. The use of the SSI technique on these soft organs/tissues requires a systematic investigation on the salient feature of the ECE in *anisotropic* soft media.

**Second**, by modeling soft tissues as incompressible transversely isotropic (TI) materials, a number of authors (Gennisson et al, 2012; Gennisson et al., 2010) have recently made efforts to determine the anisotropic elastic properties of kidney and skeletal muscles using the SSI technique, but without exploring the ECE. It has thus been recognized that three constitutive parameters are required to describe an incompressible TI solid; generally these are the transverse and longitudinal shear moduli and the elastic modulus. The studies above assessed the shear moduli $\mu_T$ and $\mu_L$, but no attempt was made to evaluate the *elastic modulus* $E_L$ using SSI technique.

**Third**, the contact between the probe and the tissue can lead to a deformation of the medium. In that respect, it is necessary and important to examine the propagation of the shear wave generated by the moving source in a *deformed* anisotropic soft tissue. The resulting study of this acousto-elastic effect will, on the one hand, enable us to quantitatively evaluate the effects of finite deformation on the determination of anisotropic properties of soft tissues using the SSI technique. On the other hand, it will also be useful to develop an inverse approach in order to determine the *in vivo* hyperelastic properties of an anisotropic soft tissue.

Bearing the above issues in mind, we investigate the ECE in an anisotropic soft material, and propose an inverse approach to infer both the linear anisotropic elastic parameters and the hyperelastic parameters of the material. Part I of the paper is organized as follows. Section 2 presents the elastodynamic model describing the elastic waves generated in an anisotropic elastic medium by a pulse load and a uniformly moving point force. We give the dispersion relationship of the TI elastic medium, which can be used to determine the correlation between the phase velocities and the group velocities of the shear waves induced by the moving point force. We carry out a theoretical analysis in Section 3 to derive the displacement field caused by a moving point force in both isotropic elastic solids and a special anisotropic elastic medium. The results help understand and interpret the ECE, at least in these cases. We perform numerical simulations in Section 4 to reveal the salient features of the ECE in more general cases. In Section 5, we provide a theoretical analysis based on the incremental elastodynamic theory (Ogden, 2007), to investigate the influence of a finite deformation on the propagation of shear waves in an anisotropic solid. Based on the analysis conducted in Sections 3-5, we propose an inverse method in Section 6 to determine the linear and hyperelastic parameters of anisotropic soft materials using the SSI technique. We conduct both theoretical analysis and numerical experiments to validate the proposed inverse method. Finally we provide some concluding remarks in Section 7.

In Part II of the paper, phantom experiments are conducted to further verify the proposed method. *Ex vivo* and *in vivo* experiments are also carried out to demonstrate the usefulness of the novel method in practice.

## 2 Wave motion in linear transversely anisotropic elastic media

In this section, we focus on the linear incompressible TI material model. It has three independent material constants (Spencer, 1984; Chadwick, 1993; Destrade et al. 2002, Rouze et al. 2013; also see the Supporting Information (SI) 1), taken here as $\mu_L$ and $E_L$, the initial shear modulus and extension modulus along the fiber direction, respectively, and $\mu_T$, the shear modulus perpendicular to that direction. In this section and henceforth, we choose a Cartesian coordinate system is such that its $x_3$-axis is aligned with the fibers and $x_1$ and $x_2$ are arbitrarily chosen in the perpendicular plane, see Fig 2. In this coordinate system, the relationships between the elastic stiffnesses $c_{ijkl}$ and $\mu_L$, $\mu_T$ and $E_L$ are straightforward to derive, see SI1.

We consider the propagation of homogeneous plane waves in the form

$$u_i(\mathbf{x},t) = U_i \exp\left[i(\mathbf{k}\cdot\mathbf{x} - \omega t)\right], \tag{2.1}$$

where $k = |\mathbf{k}|$ is the wave number, $\mathbf{k} = k\hat{\mathbf{k}}$ is the wave vector, $\hat{\mathbf{k}}$ is the unit vector in the direction of propagation, $\omega$ is the frequency, and **U** is the amplitude (or polarization) vector. Chadwick (1989) established that two transverse waves may propagate in the medium (see also Papazoglou et al., 2006; Rouze et al., 2013; see also SI1), with speeds given by

$$\begin{cases} \rho v_{SH}^2 = \mu_T \left( \hat{k}_1^2 + \hat{k}_2^2 \right) + \mu_L \hat{k}_3^2 \\ \rho v_{qSV}^2 = \mu_L + \left( E_L + \mu_T - 4\mu_L \right)\left(1 - \hat{k}_3^2 \right)\hat{k}_3^2 \end{cases}, \quad (2.2)$$

where we follow the geophysics convention for the subscripts 'SH' and 'SV' (Thomsen, 1986). The first (Shear Horizontal) wave is a pure transverse mode, polarized along **M** × **k** and the second (quasi-Shear Vertical) is polarized along **M** × **k** × **k**.

The two shear modes may be involved simultaneously in the use of SSI technique. In previous studies (Gennisson et al., 2010; Lee et al., 2012; Eby et al., 2015), the axis of the ultrasound probe is perpendicular to the fibers, so that the resulting shear waves are the SH modes according to the definition above. Rotating the probe about its axis (see Fig. 2b) changes the shear wave speeds continuously, giving access to the linear anisotropic properties of muscles for example (Gennisson et al., 2010; Lee et al., 2012). However, we can see from Eq.(2.2) that the phase velocities of the SH mode depend only on the two elastic parameters $\mu_T$ and $\mu_L$, and that the protocol does not yield the parameter $E_L$. Bearing this issue in mind we will follow Rouze et al. (2013) to propose a method to determine $E_L$ by using the qSV mode. The key idea is shown in Fig. 2c. Indeed, when the axis of the ultrasound probe is not perpendicular to the fibers, neither are the polarization directions of the resulting shear waves. In this case, the shear wave of qSV mode can be evaluated using the SSI technique, which can be related to the parameter $E_L$.

To illustrate the dependence of shear wave speeds of the qSV mode on the elastic parameter $E_L$, we introduce the following parameter $C$

$$C = \frac{\left( E_L + m_T - 4m_L \right)}{2}, \quad (2.3)$$

so that Eq.(2.2) can be rewritten as

$$\rho v_{qSV}^2 = \mu_L + 2C \left( 1 - \hat{k}_3^2 \right)\hat{k}_3^2. \quad (2.3)$$

In Fig 3, the spatial distributions of the phase velocities of the qSV mode shear waves are plotted for different propagation directions. It can be seen that the distribution of the wave speeds depends strongly on the parameter $C$. In general, Eq. (2.3) indicates that it is possible to infer $E_L$ from $v_{qSV}$ once $\mu_T$ and $\mu_L$ are known form the measurements of $v_{SH}$.

## 3  Theoretical analysis of the Elastic Cherenkov Effect

The ECE forms the theoretical basis of the SSI technique. So far, both a theoretical treatment and experimental evidence of the ECE have been reported in isotropic elastic solids (Bercoff et al., 2004a; 2004b), but its existence and key features in anisotropic elastic solids have not been fully revealed yet. In this section, we first briefly revisit the ECE in isotropic elastic solids. Although the results can be also obtained from the original theory by Eason et al. (1955), here we base our derivations on the more recent treatments by Dowling and Williams (1983) and Bercoff et al. (2004b). Second, we derive some theoretical and analytical solutions for the ECE in a special kind of anisotropic elastic solids, namely incompressible TI solids with $C=0$. In Section 4 we will perform FE simulations to investigate the ECE in more general TI soft media where $C \neq 0$.

### 3.1  The ECE in an isotropic elastic soft media

The displacement field induced by a variable body force $f_i$ depends on time $t$ and position $\mathbf{x}$ via the following integral

$$u_i(\mathbf{x},t) = \iint f_m(\boldsymbol{\xi},\tau) G_{im}(\mathbf{x}-\boldsymbol{\xi}, t-\tau) \, d\tau d\boldsymbol{\xi}. \tag{3.1}$$

where $G_{im}$ is the Green function (see Achenbach (1973) or Auld (1990) for details).

The uniformly moving point force is imposed along the same direction as the propagation direction and may be written in the following form

$$f_i(\mathbf{x},t) = a_i \delta(\mathbf{x} - v_e t \mathbf{a}), \tag{3.2}$$

where $\mathbf{a}$ is the unit vector along the moving direction and $v_e$ is the moving speed. For an isotropic elastic solid with Lamé constants as $\lambda$ and $\mu$, the *Green function* is given by (Aki and Richards, 2002)

$$G_{im}^{Iso}(\mathbf{x},t) = \frac{1}{4\pi\rho}(3\gamma_i\gamma_m - \delta_{im})\frac{1}{r^3}\int_{r/v_p}^{r/v_s}\tau\delta(t-\tau)\mathrm{d}\tau$$
$$+ \frac{1}{4\pi\rho v_p^2}\gamma_i\gamma_m\frac{1}{r}\delta\left(t-\frac{r}{v_p}\right) + \frac{1}{4\pi\rho v_s^2}(\delta_{im}-\gamma_i\gamma_m)\frac{1}{r}\delta\left(t-\frac{r}{v_s}\right). \quad (3.3)$$

where $r=|\mathbf{x}|$, $\gamma_i = \frac{x_i}{r} = \frac{\partial r}{\partial x_i}$, and $v_s = \sqrt{\mu/\rho}$, $v_p = \sqrt{(\lambda+2\mu)/\rho}$ are the velocities of the $S$ wave (shear wave, secondary wave) and $P$ wave (pressure wave, primary wave) in isotropic elastic media, respectively. The superscript 'Iso' denotes isotropic materials. Neglecting the coupling components (the first term in Eq.(3.3)), as they decay quickly away from the sources (Aki and Richards, 2002), and the $P$ wave term (the second term in Eq.(3.3)) because $v_p \gg v_s$ in an incompressible soft solid, Eq.(3.3) can be approximately written as

$$G_{im}^{Iso} \approx \frac{1}{4\pi\rho v_s^2}(\delta_{im}-\gamma_i\gamma_m)\frac{1}{r}\delta\left(t-\frac{r}{v_s}\right). \quad (3.4)$$

Without losing generality, let $\mathbf{a}=(0,0,1)$, which indicates that the force is moving along $x_3$. Inserting Eqs.(3.4) and (3.2) into Eq.(3.1) (see SI2 for the detailed derivations), we have

$$u_i(\mathbf{x},t) = \frac{1}{\sqrt{1-M_{Iso}^2\sin^2\Theta(t)}} F^{Iso}\left(\mathbf{x},t,\tau_1^{Iso},\tau_2^{Iso}\right), \quad (3.5)$$

where $M_{Iso} = v_e/v_s$ is defined as the *Mach number*,

$$\tau_{1,2}^{Iso} = t + \frac{R(t)}{v_s(M_{Iso}^2-1)}\left(M_{Iso}\cos\Theta(t)\pm\sqrt{1-M_{Iso}^2\sin^2\Theta(t)}\right), \quad (3.6)$$

$$F^{Iso}\left(\mathbf{x},t,\tau_1^{Iso},\tau_2^{Iso}\right) = \sum_{k=1}^{2}\frac{1}{4\pi\mu R(t)}\left(\delta_{i3} - \frac{\partial R(\tau_k^{Iso})}{\partial x_i}\frac{\partial R(\tau_k^{Iso})}{\partial x_3}\right). \quad (3.7)$$

and

$$\begin{cases} \mathbf{R}(t) = \mathbf{x} - v_e t\mathbf{a} \\ \Theta(t) = \arccos\left(\frac{\mathbf{R}\times\mathbf{a}}{R}\right) \end{cases}, \quad (3.8)$$

$R(t)=|\mathbf{R}(t)|$ is the distance between a spatial point $\mathbf{x}$ and the point where the moving force is applied at time $t$, and $\Theta(t)$ is the angle between the vector $\mathbf{a}$ and

$\mathbf{R}(t)$, see Fig. 4a.

From Eq.(3.6), we see that the following inequality must hold

$$1 - M_{Iso}^2 \sin^2 \Theta(t) \geq 0. \tag{3.9}$$

In the supersonic case $M_{iso} > 1$, where the speed of the moving point force is greater than the velocities of the resulting shear waves, we thus have

$$|\sin \Theta| \leq \frac{1}{M_{iso}}, \tag{3.10}$$

which indicates that the displacements are confined in a *Mach cone*. Furthermore, Eq.(3.5) shows that the displacements are singular on the cone surface. This is the ECE phenomenon, demonstrated experimentally by Bercoff et al. (2004b).

**3.2 The ECE in special incompressible TI soft media with $C = 0$**

A number of authors have investigated the response of an anisotropic elastic solid subjected to moving line forces (Stroh, 1962; Asaro et al., 1973; Ting, 1996; Wu, 2002; Iovane et al., 2004; 2005). In this paper, we perform an analytical study on wave motion in an anisotropic elastic soft media by considering the moving point force in order to reveal the key features of the ECE in an anisotropic elastic solid.

It remains a challenge to obtain the Green function in Eq.(3.1) in analytical form for a general anisotropic elastic solid; however, it is possible to obtain $G_{im}(\mathbf{x},t)$ in analytical form for some special kinds of anisotropic elastic solids. For instance, for incompressible TI solids with $C = 0$, the *Green function* has been obtained through the high-order ray theory (Vavryčuk, 2001). Under the condition of the incompressibility, and neglecting the coupling components, the solution is

$$G_{im}^{TI}(\mathbf{x},t) \approx \frac{1}{4\pi \rho^{-1/2}} \left\{ \frac{1}{\sqrt{\mu_L^3}} \frac{g_{2i} g_{2m}}{\tau_2} \delta(t - \tau_2) + \frac{1}{\mu_T \sqrt{\mu_L}} \frac{g_{3i} g_{3m}}{\tau_3} \delta(t - \tau_3) \right\} \tag{3.11}$$

where

$$\begin{cases} \mathbf{g}_2 = \frac{-1}{\sqrt{\gamma_1^2 + \gamma_2^2}} \left( -\gamma_2 \gamma_3, -\gamma_2 \gamma_3, \gamma_1^2 + \gamma_2^2 \right) \\ \mathbf{g}_3 = \frac{1}{\sqrt{\gamma_1^2 + \gamma_2^2}} \left( \gamma_2, -\gamma_1 \right) \end{cases}, \tag{3.12}$$

and

$$\begin{cases} \tau_2 = \dfrac{r}{\sqrt{\mu_L/\rho}} \\ \tau_3 = \dfrac{r}{\sqrt{\mu_T/\rho}}\sqrt{\gamma_1^2 + \gamma_2^2 + \dfrac{\mu_T}{\mu_L}\gamma_3^2} \end{cases}. \tag{3.13}$$

Using Eq.(3.11), the displacement field induced by the moving point force for the case of $C=0$ can be derived. The key results are given below and the derivations can be found in SI3.

For a TI media, without loss of generality, we let the moving direction of the source $\mathbf{a}$ lie in $x_1 - x_3$ plane and $\alpha$ be the angle between $x_3$-axis and $\mathbf{a}$, so that $\mathbf{a} = (-\sin\alpha, 0, \cos\alpha)$ as shown in Fig. 4b. The two transverse wave modes, SH and qSV, correspond to the successive two terms in the right hand side of Eq.(3.11), respectively. By separating the resulting displacements into two parts $u_i = u_i^{qSV} + u_i^{SH}$, say, we have

$$u_i^{qSV} = \dfrac{1}{\sqrt{1 - M_{qSV}^2 \sin^2\Theta(t)}} F^{qSV}\left(\mathbf{x}, t, \tau_1^{qSV}, \tau_2^{qSV}\right), \tag{3.14}$$

where $M_{qSV} = v_e/v_{qSV}$ and $v_{qSV} = \sqrt{\mu_L/\rho}$ according to Eq.(2.4) when $C=0$,

$$\tau_{1,2}^{qSV} = t + \dfrac{R(t)}{(M_{qSV}^2 - 1)\sqrt{\mu_L/\rho}}\left(M_{qSV}\cos\Theta(t) \pm \sqrt{1 - M_{qSV}^2 \sin^2\Theta(t)}\right), \tag{3.15}$$

and the definitions of $\mathbf{R}(t)$ and $\Theta(t)$ are given by Eq.**Error! Reference source not found.**.

Eq.(3.14) is basically in the same form as Eq.(3.5), as expected because the phase velocity $v_{qSV}$ and the group velocity for the qSV mode are independent of the spatial directions for $C=0$.

For the SH mode, the phase velocity $v_{SH}$ varies with spatial directions; in this case the resulting Mach cone is no longer a perfect cone (Fig. 4b). In order to make the expression of the resulting displacement more concise and consistent with Eq.(3.5) or Eq.(3.14), we introduce the coordinate transformation

$$\begin{cases} x_1' = x_1 \\ x_2' = x_2 \\ x_3' = \beta x_3 \end{cases}, \tag{3.16}$$

where $\beta^2 = \frac{\mu_T}{\mu_L}$ is a dimensionless parameter which gives a measure of the anisotropic properties of the material. When $C=0$ and $\beta=1$, the material is isotropic; while when $C \neq 0$ and $\beta=1$, the phase velocities of the SH mode do not vary with spatial directions based on Eq.(2.4), although the material may be anisotropic. In the new coordinates system, the speed $v'_e$ for the moving source is

$$v'_e = v_e \sqrt{\sin^2 a + b^2 \cos^2 a}, \tag{3.17}$$

and the unit vector in the moving direction is

$$\mathbf{a}' = \frac{(\sin\alpha, 0, \beta\cos\alpha)}{\sqrt{\sin^2\alpha + \beta^2 \cos^2\alpha}}. \tag{3.18}$$

The resulting displacement for the SH mode in the transformed coordinate system is (see SI4 for detail)

$$u_i^{SH} = \frac{1}{\sqrt{1 - M_{SH}^2 \sin^2 \Theta_{SH}}} F^{SH}\left(\mathbf{x}', t, \tau_1^{SH}, \tau_2^{SH}\right), \tag{3.19}$$

where $M_{SH} = v'_e / \sqrt{\mu_T/\rho}$ is defined as the Mach number,

$$\tau_{1,2}^{SH} = t + \frac{R'}{\sqrt{\mu_T/\rho}\left(M_{SH}^2 - 1\right)} \left(M_{SH} \cos\Theta'(t) \pm \sqrt{1 - M_{SH}^2 \sin\Theta'(t)}\right), \tag{3.20}$$

and

$$\begin{cases} \mathbf{R}'(t) = \mathbf{x}' - v'_e t \mathbf{a}' \\ \cos\Theta'(t) = \arccos\left(\frac{\mathbf{R}' \cdot \mathbf{a}'}{R'}\right) \end{cases}. \tag{3.21}$$

Eq.(3.19) shows that the displacement is confined to a Mach cone in the transformed coordinate system as illustrated in Fig. 5. This transformation provides a useful tool to study the profile of the wave fronts in an anisotropic soft media generated by a moving source, e.g., the existence of the Mach cone and its cone angle. In the SSI technique (Bercoff et al., 2004a), the source moves with supersonic speed and the Mach number is very large. In that case, our analytical results given by Eq. (3.18) below reveal that quasi-plane waves can be generated. The wave speeds can be measured in experiments by relying on the time of flight algorithm (McLaughlin and Renzi, 2006; Tanter et al., 2008), which can be used to further infer the elastic properties of anisotropic soft materials.

The analytic solution was derived here under the condition $C=0$. It has been illustrated in Fig. 3 that the parameter $C$ plays an important role in determining the spatial variation of the shear wave velocities. We now investigate the ECE in more general cases, when $C \neq 0$.

## 4 Computational studies on the ECE in an incompressible TI solid

For incompressible TI solids with $C \neq 0$, it is difficult to obtain the solutions in analytical forms similar to Eqs. (3.13) and (3.19). In this section we use finite element analysis (FEA) to study the ECE in more general cases.

The FEA is conducted using the commercial software Abaqus/explicit (2010). A cubic solid of size $40\text{mm} \times 40\text{mm} \times 40\text{mm}$ is used to model the soft media. The model contains 4,096,000 C3D8R elements. Following previous studies on acoustic radiation force (Palmeri et al., 2005; Rouze et al., 2013), the moving source is modeled as the following body force with three-dimensional Gaussian distribution, traveling uniformly with speed $v_e$,

$$f_i(\mathbf{x},t) = a_i F_0 e^{-\frac{(\mathbf{x}-v_e t \mathbf{a})^2}{r_0^2}}, \qquad (4.1)$$

where $F_0 = 10^{-3} N$ denotes the amplitude of the force, the unit vector $\mathbf{a}$ is along the moving direction, and $r_0 = 0.5\text{mm}$. We rely on the user subroutine of VDload in Abaqus.

As mentioned in Section 2, the parameter $C$ plays a key role in investigating the ECE in an anisotropic soft material. We separate the class of incompressible TI solids into three categories: $C > 0$, $C = 0$ and $C < 0$. In our FEA we take three different TI materials with the representative material constants given in Table 1.

Table 1. Elastic parameters used in the FEA models

|  | $E_L$ ( kPa) | $\mu_L$ ( kPa) | $\mu_T$ ( kPa) |
|---|---|---|---|
| C= 62.5 kPa | 216 | 25 | 9 |
| C= 0 | 91 | 25 | 9 |
| C= -21.875 kPa | 47.25 | 25 | 9 |

We again choose a coordinate system such that $x_3$-axis is parallel to the fibers. The source moves in a line making an angle $\alpha$ with the fibers, so that $\mathbf{a} = (-\sin\alpha, 0, \cos\alpha)$. We investigate the following three different cases in turn:

$a = 0°, 90°, 45°$.

### 4.1 Source moving along the fibers $\alpha = 0°$

When the source moves along the direction $\alpha = 0°$, the problem is axisymmetric. In this case, the shear wave of qSV mode is the main concern, and its speed can be measured with the SSI technique.

We first study the case where the Mach number, $M_{qSV} = v_e / \sqrt{\mu_L / \rho}$, is relatively small, in order to reveal the key feature of the ECE in an anisotropic soft medium. The source speed is taken as $v_e = 15 \text{ m/s}$ in our simulations, which corresponds to $M_{qSV} = 3$. The representative results shown in Fig.7 indicate that the resulting shear-wave Mach cones depend strongly on the parameter $C$. For instance for $C > 0$ (Figs. 7a-d), we first measure the angle of Mach cone (Fig. 7a) based on the finite element results. On the other hand, inserting the wave vector of $\hat{\mathbf{k}} = (\sin\theta, 0, \cos\theta)$ into Eq.(2.2) (so that the angle of the Mach cone is $(90° - \theta)$), and using the geometric condition $v_{qSV} / v_e = \sin\theta$, we have

$$\frac{\sqrt{(m_L + 2C \sin^2 q \cos^2 q)/r}}{v_e} = \sin(90° - q) = \cos(q). \tag{4.2}$$

Solving Eq.(4.2) with the values of the first line in Table 1 gives $\theta = 25.5°$, in good agreement with the value obtained by the FEA (Fig. 7a).

Fig. 6 shows that the shear-wave Mach cones are dependent on the parameter $C$ (or the elastic parameter $E_L$), and in this sense it is possible in theory to evaluate the elastic parameter $E_L$ by measuring the angle of the Mach cone. However, our analysis shows that the variation of the cone angle is not sensitive to the variation in $E_L$, indicating that the solution to the inverse problem will be very sensitive to data errors.

Fig. 6 also shows that the direction of the wave vector $\hat{\mathbf{k}}$ is not always consistent with the direction of the group velocity $\mathbf{v}_g$, as expected for wave motion in an anisotropic media (Thomsen, 1986). A schematic of this phenomenon is given in Fig. 6. For all three cases of Table 1, the moving distances of the source are the same but the interfered wave fronts are different, simply because the directions of the group velocities are not the same (see Figs.6d, 6h and 6l). In general, the interfered wave fronts in anisotropic media travel along the directions of group velocities, which are usually different from the directions of the wave vectors, except in some special cases (e.g. $C=0$ in the present problem).

We then investigate the case where the Mach number is large. In experiments (Bercoff et al., 2004b), the source can move with supersonic speed, and the Mach number can be very large (~1,000). We can see from Eq.(4.2) that the angle of the Mach cone, $(90°-\theta)$, is close to zero when $M_{qSV} = v_e / \sqrt{\mu_L/\rho}$ is large. In our FEA model, we take $M_{qSV} = 30$. From Figs. 7a, 7d and 7g, we see that the resulting angles of the Mach cone for all the three cases are very close to zero, i.e., $\theta \approx 90°$. In this case, the wave vector is almost aligned with $x_1$: $\hat{\mathbf{k}} = (1,0,0)$. In practical experiments, the wave speed $v_{qSV,\perp}$ along the direction of wave vector is measured and the phase velocity formula (Eq.**Error! Reference source not found.**) can be simply used to determine $\mu_L$

$$\rho v_{qSV,\perp}^2 = \mu_L, \tag{4.3}$$

where the subscript '$\perp$' denotes that the propagation direction of the shear wave is perpendicular to the fibers. In our analysis, the parameter $\mu_L$ is taken as constant in the three cases; therefore, the phase velocities of the shear waves should also have the same values (Fig. 7).

We also plot the normalized displacements in Fig. 7j for two points located along the direction of the wave vector (i.e., along the $x_2$-axis). Then, by tracking the time delay of the peaks on the curves, we can calculate the speed of the shear wave $v_{qSV,\perp}$.

**4.2 Source moving perpendicular to the fibers $\alpha = 90°$**

In this case, only the SH modes exist in the ultrasound imaging surface according to the discussion in Section 2. The Mach number is $M_{SH} = v'_e / \sqrt{\mu_T/\rho}$, where $v'_e$ defined in Eq.(3.17) equals $v_e$ when $\alpha = 90°$. In our simulations, we took $M_{SH} = 50$. The results are given in Figs. 8a-c. They show that the resulting shear waves are basically the same even though the elastic parameter $C$ (or $E_L$) is different in the three cases, which confirms the analytical result that the SH mode is only dependent on $\mu_L$ and $\mu_T$ as mentioned in Section 3. In our FEA, we fixed the parameters $\mu_L$ and $\mu_T$ were fixed according to Table 1, which explains why the speeds of the resulting waves for the three cases are basically the same. The shear wave speeds $v_{SH,\perp}$ and $v_{SH,\parallel}$ perpendicular and along the fibers, respectively, are determined from the variation of the displacements with time at four characteristic points as plotted in Fig. 8j. Considering that the quasi-plane waves are generated at a high Mach number, the correlations between the phase velocities and the shear moduli along and perpendicular to the fibers are given by (Thomsen, 1986)

$$\begin{cases} \rho v_{SH,\perp}^2 = \mu_T \\ \rho v_{SH,\parallel}^2 = \mu_L \end{cases}. \qquad (3.1)$$

A number of authors used Eqs. (3.4) together with the SSI technique (Gennisson et al., 2010; Lee et al., 2012) to determine $\mu_L$ and $\mu_T$ for an incompressible TI soft tissue.

Our analysis above shows that Eq. (4.4) is applicable provided that the moving speed of the source is high and hence the Mach number is large, say greater than 30.

In Section 3 we introduced a change of coordinates to obtain the Mach cone of the SH mode in the form of a perfectly circular cone. As a further illustration, a comparison between the Mach cone in the original coordinate and the one in transformed coordinate system is given in Fig. 9 by taking $C = 62.5$ kPa as an example. It is clear that in the transformed coordinate system, the Mach cone in the form of circular cone is achieved, as expected.

### 4.3 Source moving oblique to the fibers $\alpha = 45°$

The analysis above shows that in the cases of $\alpha = 90°$ and $\alpha = 0°$, shear-wave Mach cones can be formed to produce a reliable means to determine the initial shear moduli $\mu_L$ and $\mu_T$ from the velocities of the quasi-plane waves generated by the moving shear source, simply by using Eq. (4.4). However, it remains a challenge to determine the third elastic parameter $E_L$. Bearing this important issue in mind, we investigate the case where the angle between the moving direction of the source and the material symmetric axis is $45°$ (Rouze et al., 2013), see Fig. 2c. In this case, the qSV mode may be used to access the parameter $E_L$.

Fig. 10 gives the computational results. We clearly see that the resulting shear-wave Mach cones are significantly different from those observed in the cases of $\alpha = 90°$ and $\alpha = 0°$, and they strongly depend on the parameter $C$ (or $E_L$). When $C > 0$ (Fig. 10g), two shear waves with different polarization directions are clearly observed, in contrast to the cases where $C \leq 0$ (Figs. 10h and 10i). This may be explained as follows. When $C > 0$, two wave modes are excited and they have different speeds in the ultrasound imaging surface as illustrated in Fig. 2c. Then Eq.**Error! Reference source not found.** and the definition of parameter $C$ indicate that the phase velocity of the qSV mode is greater than that of the other mode. When $C \leq 0$, the displacements generated by the source moving at high speed will be approximately parallel to the loading direction and therefore only the qSV mode can be observed in the ultrasound imaging surface as shown in Fig. 10. In summary, the interfered wave fronts tracked in experiments using SSI technique (Bercoff et al., 2004a) will always be the qSV modes when $\alpha = 45°$. The normalized displacements perpendicular to the moving direction of the source for two characteristic points are plotted in Figs. 10g-i, enabling us to evaluate the speeds of the quasi-plane shear waves. Furthermore, the following correlation between the phase velocity of the qSV mode and the material parameters may be used establish an inverse approach to measure the parameter $C$ (or $E_L$) as shown in detail in Section 6.

$$\rho v_{qSV,45}^2 = \mu_L + C/2. \tag{4.5}$$

## 5 Propagation of shear waves in a deformed anisotropic soft material

In the previous sections, we investigated the ECE in an anisotropic, incompressible, linear (undeformed, unstressed) material. The results may be used together with the dynamic elastography method, e.g. SSI technique (Bercoff et al., 2004a), to determine the anisotropic elastic parameters of biological soft tissues. It is worth noticing that *(a)* in clinical use, the contact of the probe with tissue may lead to finite deformation in the soft tissue and *(b)* knowing how shear waves propagate in a deformed soft tissue (the theory of acousto-elasticity) gives access to its hyperelastic (nonlinear) properties (Gennisson et al, 2007; Latorre-Ossa et al., 2012; Jiang et al, 2015a; 2015b). In this section we thus investigate the propagation of shear waves in a deformed incompressible TI soft media.

The theory of elastic wave propagation in deformed solids can be dated back to the works of Hadamard (1903), Brillouin (1925) and Biot (1940). It was later rewritten in compact form by Ogden and collaborators (see Ogden (2007) for a review or Ogden and Singh (2011)) and further developed within the framework of finite elasticity. Here we rely on these equations to derive analytical solutions predicting the speed of the shear waves generated in an incompressible anisotropic soft tissue by a moving source. In our derivations, we use the constitutive relationship proposed by Murphy (2013), as it is compatible with the linear elastic TI model at infinitesimal deformation and motions. This choice will provide a strong link with ultrasonic measurements made on un-deformed solids, but any other hyperelastic TI model can equally be used in our analysis.

## 5.1 Governing equations

Here we briefly recall the governing equations used to analyze the propagation of homogeneous shear waves in a deformed hyperelastic solid, see Ogden (2007) for more details.

### 5.1.1 Equation of motion

We consider small-amplitude motions in an incompressible TI material with strain energy density $W$, which has been subject to a large homogeneous deformation described by the constant deformation gradient $\mathbf{F} = \partial \mathbf{x} / \partial \mathbf{X}$ where $\mathbf{x}$ and $\mathbf{X}$ are the spatial and material coordinates, respectively. The displacement $\mathbf{u}(\mathbf{x},t)$ satisfies the following incremental equations of motion and of incompressibility (Ogden, 2007)

$$A_{0piqj} u_{j,qp} - \delta p_{,i} = \rho \ddot{u}_i, \qquad u_{i,i} = 0, \qquad (5.1)$$

where $\delta p$ is the increment of the Lagrange multiplier $p$ due to the internal constraint of incompressibility and $\mathbf{A_0}$ is the fourth-order tensor of instantaneous elastic moduli, with components

$$A_{0piqj} = F_{pa} F_{qb} \frac{\partial^2 W}{\partial F_{ia} \partial F_{jb}}. \qquad (5.2)$$

## 5.2 Constitutive relation

A good number of constitutive models have been proposed over the years to characterize incompressible TI soft tissues (e.g. Humphrey and Yin, 1987; Merodio and Ogden 2003; 2005; Destrade et al., 2013; Murphy, 2013). In general, their strain-energy function $W$ may be written as a function of four invariants,

$$W = W(I_1, I_2, I_4, I_5), \qquad (5.3)$$

say, where $I_1 = tr\mathbf{C}$, $I_2 = \frac{1}{2}[(tr\mathbf{C})^2 - tr(\mathbf{C}^2)]$ are the isotropic invariants, $\mathbf{C} = \mathbf{F}^T\mathbf{F}$ is the right Cauchy-Green deformation tensor, and $I_4$ and $I_5$ are the anisotropic invariants,

$$I_4 = \mathbf{MCM}, \quad I_5 = \mathbf{MC^2M} \tag{5.4}$$

Here the unit vector $\mathbf{M}$ is along the fibers, which we chose to be aligned with the $x_3$-axis of the coordinate system, so that $\mathbf{M} = (0,0,1)$.

The corresponding Cauchy stress tensor is (Spencer, 1972)

$$\begin{aligned}\boldsymbol{\sigma} = &-p\mathbf{I} + 2W_1\mathbf{B} + 2W_2(I_1\mathbf{B}-\mathbf{B}^2) + 2W_4\mathbf{FM}\otimes\mathbf{FM} \\ &+ 2W_5(\mathbf{FM}\otimes\mathbf{BFM}+\mathbf{BFM}\otimes\mathbf{FM})\end{aligned}, \tag{5.5}$$

where $\mathbf{B} = \mathbf{FF}^T$ is the left Cauchy-Green deformation tensor and $W_i = \dfrac{\partial W}{\partial I_i}$ $(i \in \{1,2,4,5\})$. When the material is un-deformed ($\mathbf{F} = \mathbf{I}$) it is assumed to be stress-free, so that

$$p^0 = 2W_1^0 + 4W_2^0, \quad W_4^0 + 2W_5^0 = 0 \tag{5.6}$$

where the superscript 0 indicates that the quantities are evaluated in the ground state, where $I_1 = I_2 = 3$, $I_4 = I_5 = 1$.

For infinitesimal deformations, three independent elastic parameters, i.e., $\mu_T$, $\mu_L$ and $E_L$ are required to describe the mechanical behavior of the incompressible TI solid, as recalled earlier. Merodio and Ogden (2003, 2005) presented the conditions to ensure the compatibility between the linear elastic and hyperelastic models; they can be written as (Murphy, 2013)

$$\begin{cases} 2W_1^0 + 2W_2^0 = \mu_T \\ 2W_1^0 + 2W_2^0 + 2W_5^0 = \mu_L \\ 4W_{44}^0 + 16W_{45}^0 + 16W_{55}^0 = E_L + \mu_T - 4\mu_L \end{cases}, \tag{5.7}$$

where $W_{ij}$ is $\dfrac{\partial^2 W}{\partial I_i \partial I_j}$ $(i,j \in \{1,2,4,5\})$.

Murphy (2013) pointed out that $W$ should include both $I_4$ and $I_5$ in order to meet the initial condition given by Eq. (5.7). Furthermore, the strain-energy function may be written in the following form

$$W = F(I_1, I_4) + \frac{\mu_T - \mu_L}{2}(2I_4 - I_5 - 1). \tag{5.8}$$

It can be verified that Eq.(5.8) is compatible with the linear elastic model. Bearing Eq.

(5.23) in mind, Murphy (2013) generalized the material model proposed by Humphrey-Yin (1987) to the following form, which is used in this study

$$W = \frac{m_T}{2c_2}[e^{c_2(I_1-3)} - 1] + \frac{E_L + m_T - 4m_L}{2c_4}[e^{c_4(I_4^{1/2}-1)^2} - 1] + \frac{m_T - m_L}{2}(2I_4 - I_5 - 1), \quad (5.9)$$

where $c_2 > 0$ and $c_4 > 0$ are isotropic and anisotropic strain hardening parameters, respectively, and $I_4 = \mathbf{M} \cdot \mathbf{CM}$ is the squared stretch in the direction of the fibers. To simplify expressions, we will use $A$, $B$ and $C$ to denote $\frac{\mu_T}{2}$, $\frac{\mu_T - \mu_L}{2}$ and $\frac{E_L + \mu_T - 4\mu_L}{2}$, respectively.

Using the strain-energy function Eq.(5.9), we can compute the elastic moduli explicitly as (e.g., see Destrade (2015)),

$$A_{0jikl} = 2W_1 \delta_{il} B_{jk} + 2W_2 \delta_{il} m_j m_k + 2W_5[\delta_{il}(\bm{Bm})_j m_k + \delta_{il}(\bm{Bm})_k m_j + B_{jk} m_i m_l + B_{ik} m_j m_l + B_{ij} m_i m_k] + 4W_{11} B_{ij} B_{kl} + 4W_{44} m_i m_j m_k m_l, \quad (5.10)$$

where $\bm{m} = \bm{BM}$ and

$$W_1 = Ae^{c_2(I_1-3)}, W_4 = 2B + Ce^{c_4(I_4^{1/2}-1)^2}(1 - I_4^{-1/2}), W_5 = -B,$$
$$W_{11} = c_2 Ae^{c_2(I_1-3)}, W_{44} = c_4 Ce^{c_4(I_4^{1/2}-1)^2}(1 - I_4^{-1/2})^2 + \frac{1}{2}Ce^{c_4(I_4^{1/2}-1)^2} I_4^{-3/2}. \quad (5.11)$$

## 5.3 Analytical solution to predict the speed of the shear wave in a deformed anisotropic soft medium

In this subsection, we study the effects of finite deformation in the soft material on the propagation of the SH wave (Fig. 2b). In Sections 3 and 4, we have shown the presence of the ECE and the quasi-plane waves generated at a high Mach number. Therefore, in this section, we consider the propagation of the plane waves of the form (Destrade et al, 2010a; 2010b)

$$\begin{cases} \mathbf{u} = \mathbf{U}e^{ik(\hat{\mathbf{k}} \cdot \mathbf{x} - vt)}, \\ \delta p = ikPe^{ik(\hat{\mathbf{k}} \cdot \mathbf{x} - vt)} \end{cases}, \quad (5.12)$$

where $\mathbf{U}$ is the amplitude of the wave (without loss of generality, $\mathbf{U}$ is taken as a unit vector), $\hat{\mathbf{k}}$ is the unit vector in the direction of wave propagation, $k$ is the wave number, $v$ is the phase velocity, and $P$ a scalar.

Inserting $\mathbf{u}$ and $\delta p$ into the incremental equations of equilibrium Eq.(5.1) gives

$$\mathbf{Q}(\hat{\mathbf{k}})\mathbf{U} - P\hat{\mathbf{k}} = \rho v^2 \mathbf{U}, \quad (5.13)$$

where $[Q(\hat{\mathbf{k}})]_{ij} = A_{0piqj}\hat{k}_p\hat{k}_q$ is the acoustical tensor. The constraint of incremental incompressibility given in Eq.(5.1) becomes

$$\mathbf{U} \cdot \hat{\mathbf{k}} = 0, \tag{5.14}$$

which means the wave is purely transverse. Taking the dot product of Eq.(5.13) with $\hat{\mathbf{k}}$ gives

$$P = \hat{\mathbf{k}} \cdot \mathbf{Q}(\hat{\mathbf{k}})\mathbf{U}. \tag{5.15}$$

Inserting Eq.(5.15) and Eq.(5.14) into Eq.(5.13), we arrive at the following symmetric eigenvalue problem

$$(\mathbf{I} - \hat{\mathbf{k}} \otimes \hat{\mathbf{k}})\mathbf{Q}(\hat{\mathbf{k}})(\mathbf{I} - \hat{\mathbf{k}} \otimes \hat{\mathbf{k}})\mathbf{U} = \rho v^2 \mathbf{U}. \tag{5.16}$$

Taking the dot product of Eq.**Error! Reference source not found.** with $\mathbf{U}$, which together with the expression of $[Q(\hat{\mathbf{k}})]_{ij}$ gives the expression of wave speed

$$\rho v^2 = \mathbf{U} \cdot \mathbf{Q}(\hat{\mathbf{k}})\mathbf{U} = A_{0piqj}\hat{k}_p\hat{k}_q U_i U_j. \tag{5.17}$$

Recall that $x_3$ is aligned with the fibers so that $\mathbf{M} = (0,0,1)$. We assume that the direction of the moving source is along $x_2$, in line with the particulars of SSI technique. In this case, the polarization direction of the SH wave is along $x_2$ as well, $\mathbf{U}=(0,1,0)$. The direction of wave propagation lies in the $x_1 - x_3$ plane and is given by $\hat{\mathbf{k}} = (\sin\theta, 0, \cos\theta)$, say.

As shown in Fig. 12, the homogeneous state of deformation can be described by

$$x_1 = \lambda_1 X_1, x_2 = \lambda_2 X_2, x_3 = \lambda_3 X_3, \tag{5.18}$$

where $\lambda_1$, $\lambda_2$ and $\lambda_3$ are the principal stretch ratios. The deformation gradient tensor $\mathbf{F}$ corresponding to this deformation state is

$$\mathbf{F} = \lambda_1 \mathbf{e}_1 \otimes \mathbf{E}_1 + \lambda_2 \mathbf{e}_2 \otimes \mathbf{E}_2 + \lambda_3 \mathbf{e}_3 \otimes \mathbf{E}_3, \tag{5.19}$$

where $\mathbf{e}_i$ and $\mathbf{E}_\alpha$ ($i, \alpha \in \{1,2,3\}$) are the base vectors of the material and spatial configurations, respectively. The constraint of incompressibility, $\det \mathbf{F} = 1$, gives $\lambda_1\lambda_2\lambda_3 = 1$. For the strain-energy function Eq.(5.9), the phase velocity $v$ is determined from Eq.(5.1) as

$$\rho v^2 = 2Ae^{c_2(I_1-3)}\lambda_1^2 \sin^2\theta + [2Ae^{c_2(I_1-3)} + 4B + 2Ce^{c_4(\lambda_3-1)^2}(1-\frac{1}{\lambda_3}) \\ - 2B(2\lambda_3^2 + \lambda_2^2)]\lambda_3^2 \cos^2\theta$$
(5.20)

where $I_1 = \lambda_1^2 + \lambda_2^2 + \lambda_3^2$.

Eq.(5.20) clearly shows how the shear wave speed depends on the material parameters and the deformation of the material. Based on Eq.(5.20), an inverse approach will be established to determine the constitutive parameters as shown in Section 6 in detail. In the absence of deformation, $\lambda_1 = \lambda_2 = \lambda_3 = 1$, and Eq.(5.20) reduces to

$$\rho v^2 = \mu_T \sin^2\theta + \mu_L \cos^2\theta,$$
(5.21)

which is consistent with Eq.**Error! Reference source not found.** when taking $\hat{\mathbf{k}} = (\sin\theta, 0, \cos\theta)$.

In the practice, the shear wave speeds in the directions along ($\theta = 0^o$) and perpendicular ($\theta = 90^o$) to the fibers are measured. Inserting $\theta = 0^o$ and $\theta = 90^o$ into the Eq.**Error! Reference source not found.**, the speeds of the shear waves in these two directions is obtained from

$$\rho v_T^2 = 2Ae^{c_2(I_1-3)}\lambda_1^2,$$
(5.22)

$$\rho v_L^2 = [2Ae^{c_2(I_1-3)} + 4B + 2Ce^{c_4(\lambda_3-1)^2}(1-\frac{1}{\lambda_3}) - 2B(2\lambda_3^2 + \lambda_2^2)]\lambda_3^2,$$
(5.23)

where we use $v_L$ and $v_T$ for the speeds along and perpendicular to the fibers, respectively.

To further illustrate the dependence of the shear wave speed given in Eqs. (5.22) and (5.23) on the deformation state, now we write $\lambda_2 = \lambda$ and $\lambda_1 = \lambda^{-\xi}$, so that $\lambda_3 = \lambda^{-(1-\xi)}$ by the constraint of incompressibility. When a uniform compression is imposed along $x_2$, the value of $\xi$ should be in the range of $0.5 < \xi < 1$, with the lower bound corresponding to uni-axial compression/equi-biaxial deformation and the upper bound to plane strain deformation. Fig. 12 gives the variation of $\rho v_T^2/\mu_T$ and $\rho v_L^2/\mu_L$ with $\xi$ for different $\lambda$s. We see that $\xi$ has significant influence only when $\lambda$ is small for the case of propagation direction perpendicular to the fibers, e.g. $\lambda = 0.75$. This conclusion agrees with that given by Jiang et al (2015a) for isotropic soft materials. However, when the propagation direction is along the material symmetric axis, $\rho v_L^2 / \mu_L$ significantly depends on $\xi$, even when the deformation is small, e.g. $\lambda = 0.9$. Besides, the greater the value of $E_L$ is, the more significant the effect of $\xi$ on $\rho v_L^2 / \mu_L$ will be. Our analytical solutions given by Eqs. (5.37) and (5.38) not only enable us to evaluate the extent to which the deformation affects the wave velocities and further the determination of anisotropic parameters but also allow us to develop an inverse approach to determine the hyperelastic parameters of anisotropic soft materials, as seen in detail in Section 6.

# 6 Inverse method to determine the linear and hyperelastic parameters of TI soft materials

In Sections 3-5, we investigated the ECE in anisotropic elastic soft media and propagation of shear waves in a deformed material. Based on the theoretical and computational results, an inverse approach will be proposed to infer the anisotropic and hyperelastic parameters of an incompressible soft material.

## 6.1 Determination of anisotropic parameters $\mu_L$, $\mu_T$ and $E_L$

Three elastic parameters $\mu_L$, $\mu_T$ and $E_L$ are required to describe a linear elastic incompressible TI soft material. The existence of the ECE in an anisotropic soft media allows us to evaluate $\mu_L$ and $\mu_T$ using the correlation between the speeds of plane shear waves and the material parameters. A number of authors have managed to measure $\mu_L$ and $\mu_T$ for skeletal muscles (Gennisson et al., 2010), skins (Luo et al., 2015 ) and kidneys (Gennisson et al., 2012) using SSI technique. However, no effort has been made to infer the parameter $E_L$ of soft tissues using the SSI technique, which may be altered by injury or disease. Besides, some caution should be taken for the determination of $\mu_L$ and $\mu_T$. For instance, Gennisson et al. (2012) recently evaluated these constants with the SSI technique by putting the ultrasound beam in turn perpendicular to, and then along the oriented renal structures. It should be pointed out that the shear speeds along these two directions are both related to the initial shear modulus $\mu_L$ according to the analysis above, and so only $\mu_L$ can be determined in that way. This is confirmed by examination of Gennisson et al.'s (2012) experimental results, revealing that the shear moduli determined using the shear waves along these two directions are indeed very close to each other. Our analysis in Section 4 shows that in the case of $\alpha = 45°$ quasi-plane waves can be formed when the Mach number is high, which enables us to determine $C$ or $E_L$ by measuring the speed of the quasi-plane waves along the direction of $\hat{\mathbf{k}} = \left(-\sin\left(\frac{\pi}{4}\right), 0, \cos\left(\frac{\pi}{4}\right)\right)$, which is denoted by $v_{qSV,45°}$.

Based on Eq. (4.5), the elastic parameter $C$ or $E_L$ can be determined using the following relations

$$C = 2\left(\rho v_{qSV,45}^2 - \mu_L\right)$$
$$E_L = 8\mu_L - \mu_T - 4\rho v_{qSV,45}^2 .$$
(6.1)

### 6.2 Determination of the hyperelastic parameter $c_2$

When the propagation direction of the shear wave is perpendicular to the fibers, we introduce the quantities $\lambda$ and $\xi$, and rewrite Eq.**Error! Reference source not found.** as follows

$$\rho v_T^2 = \mu_T e^{c_2(\lambda^2 + \lambda^{-2\xi} + \lambda^{-2(1-\xi)} - 3)} \lambda^{-2\xi}.$$
(6.2)

As explained previously, we can determine the transverse shear modulus $\mu_T$ from the measurements of $v_T$. Furthermore, the hardening parameter $c_2$ can be determined from the following equation

$$c_2 = \frac{\ln\left(\dfrac{\rho v_T^2 \lambda^{2\xi}}{\mu_T}\right)}{\lambda^2 + \lambda^{-2\xi} + \lambda^{-2(1-\xi)} - 3}.$$
(6.3)

### 6.3 Validations of the inverse method

Determining the anisotropic and hyperelastic parameters of soft tissues from speed measurements is an inverse problem. The sensitivity of the identified solutions to data errors can be assessed by introducing the condition number. The condition numbers for determination of $\mu_L$, $\mu_T$ parameter $E_L$ are equal to 2, which indicates that a 3% error in the measured wave velocity will lead to a 6% error in the identified solutions. The condition number which measures the sensitivity of the isotropic strain-hardening parameter $c_2$ to the errors in $v_T^2/\mu_T$ is

$$\text{Cond}_{c_2} = \frac{\Delta c_2}{c_2} \bigg/ \frac{\Delta(v_T^2/\mu_T)}{v_T^2/\mu_T} = \frac{1}{c_2\left(\lambda^2 + \lambda^{-2\xi} + \lambda^{-2(1-\xi)} - 3\right)}. \qquad (6.4)$$

Fig. 13 gives the plot of the condition number according to Eq. (6.4) which shows that the parameter $\lambda$ plays an important role here. In order to obtain a more reliable evaluation on the parameter $c_2$, a greater $\lambda$ should be used in practical measurements.

We then validate the inverse method using numerical experiments. In Fig. 9 d, we plot the normalized displacements of four characteristic points for $C = 62.5 \text{ kPa}$. All of the points are located in the plane perpendicular to the moving direction of the shear source. Two points are located on the $x_2$-axis (perpendicular to material symmetric axis), while the other two are located on the $x_3$-axis. The distance between each two points is $\Delta d = 2.0 \text{ mm}$. To measure the velocities of the shear waves, time delays of the peak on the curves, denoted by $\Delta t_{SH,\parallel}$ and $\Delta t_{SH,\perp}$, are measured. Then the velocities can be calculated with $v_{SH,\perp} = \frac{\Delta d}{\Delta t_{SH,\perp}}$ and $v_{SH,\parallel} = \frac{\Delta d}{\Delta t_{SH,\parallel}}$, respectively. The time delay calculated by the FEA can be read from Fig. 9 d, i.e., $\Delta t_{SH,\perp} = 0.69 \text{ ms}$ and $\Delta t_{SH,\parallel} = 0.40 \text{ ms}$, then we have $v_{SH,\perp} = 2.9 \text{ m/s}$ and $v_{SH,\parallel} = 5.0 \text{ m/s}$. From Eq. (4.4) we deduce the material parameters $\mu_T$ and $\mu_L$ as $\mu_T = 8.4 \text{ kPa}$ and $\mu_L = 25.0 \text{ kPa}$, respectively, in good agreement with the input parameters used in the FEA (Table 1).

In Fig. 10, we plot the normalized displacements for P1 and P2 along the direction of wave vector, because they are also tracked in the SSI technique. From the figures, we obtain the time delays of peaks on the curves. Furthermore, the shear wave speeds $v_{qSV,45}$ are calculated, and the parameters $C$ deduced according to Eq. (6.1). The results for the three cases are listed in Table 2. The parameters $C$ (or $E_L$), deduced from the measured speeds match those we input in the FEA well, indicating that our inverse approach is effective. *In vivo* determination of $E_L$ of soft tissues using elastography method remains a challenge and our method has great potential for practical use, as will be demonstrated in Part II of the paper.

Table 2. The measured time delays of the qSV mode in the case of $\alpha = 45°$. Shear wave speeds are calculated, from which the parameters $C$ follow.

| Input $C$ (kPa) | Time delay of the qSV mode (ms) | $v_{qSV,45°}$ (m/s) | Identified $C$ (kPa) |
|---|---|---|---|
| 62.5 | 0.262 | 7.63 | 66.4 |
| 0 | 0.408 | 4.90 | -1.98 |
| -21.875 | 0.550 | 3.64 | -23.5 |

In the determination of $\mu_T$, $\mu_L$ and $E_L$, the deformation of soft tissues caused by the contact of the ultrasound probe with soft tissues may lead to errors. Our analytical solutions given by Eqs. (5.37) and (5.38) allow us to quantitatively estimate the effects of soft tissue deformation. For instance in the determination of $\mu_T$, if we take the hardening parameter $c_2 = 3$ as an example, then a deformation of $\lambda = 0.8$ may lead to an error up to 100% in the measured $\mu_T$. In that sense, the analytical solutions proposed here provide guidelines for controlling the effects of deformation on the measurement of anisotropic parameters using SSI technique.

## 7 Concluding remarks

When the source is moving at high speed (e.g. supersonic speed), shear-wave Mach cones may be formed and quasi-plane waves generated in a soft medium. This phenomenon is named as the Elastic Ceronkov Effect (ECE) and has been studied in an isotropic soft solid, which forms the theoretical basis of the supersonic shear imaging method (Bercoff et al., 2004a; 2004b). In this paper, we investigated the ECE in an incompressible TI solid because many soft tissues belong to this class of materials. In summary the following key results have been obtained in Part I of the paper.

First, both theoretical analysis and computational studies have been performed to investigate the shear-wave Mach cones generated by the moving source. Our results clearly demonstrated that quasi-plane waves formed at high Mach numbers. In the last decades, a number of analytical solutions have been proposed to predict the speeds of shear waves for different types of anisotropic solids. These analytical solutions in the literature, together with the existence of the ECE and its salient features as revealed in this paper, enable us to get simple correlations between the shear wave speed and the material parameters. These results show that not only $\mu_T$ and $\mu_L$ but also $E_L$ may be determined from simple relations.

Second, we investigate the propagation of the shear wave in a deformed incompressible TI soft media. Based on the theory proposed by Ogden (2007) and a constitutive model proposed by Murphy (2013), we derived analytical solutions to reveal the correlation among the shear wave speeds, material parameters and the deformation of the solid.

Third, based on the theoretical solutions presented above, we proposed an inverse approach to determine the linear and the hyperelastic parameters of an incompressible TI soft material. We show that the initial shear moduli $\mu_T$ and $\mu_L$, the elastic modulus $E_L$ and the hyperelastic parameter $c_2$ can be determined using our inverse approach.

Finally, we carried out both theoretical analysis and numerical experiments to validate the inverse approach. Our theoretical analysis, introducing the concept of condition number, shows the extent to which the identified solutions are sensitive to data errors. The numerical experiments indicate that the material parameters can be determined using the proposed inverse method with good accuracy.

Figures

(a) (b)

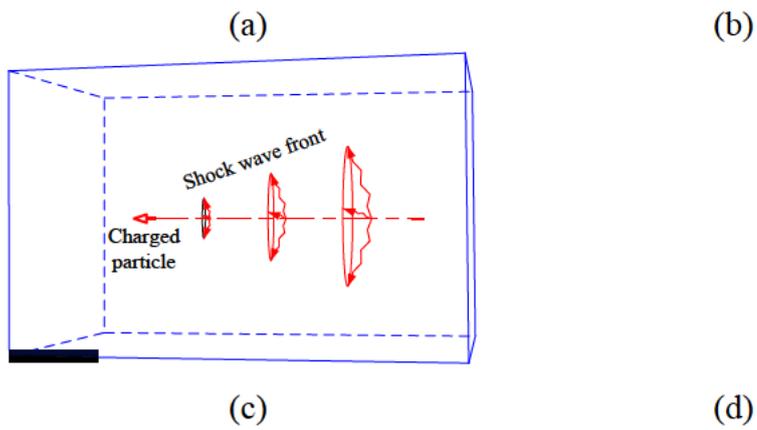

(c) (d)

Fig. 1 Four similar wave phenomenon observed in the case where the velocities of the excitation source are greater than the velocities of the resulting waves in the media. (a) "Sonic boom" phenomenon caused by a F/A-18F plane during a transonic flight; (b) the Kelvin ship-wave pattern; (c) a schematic of the Cherenkov radiation; (d) the elastic Cherenkov effect induced by the moving shear source in the soft media.

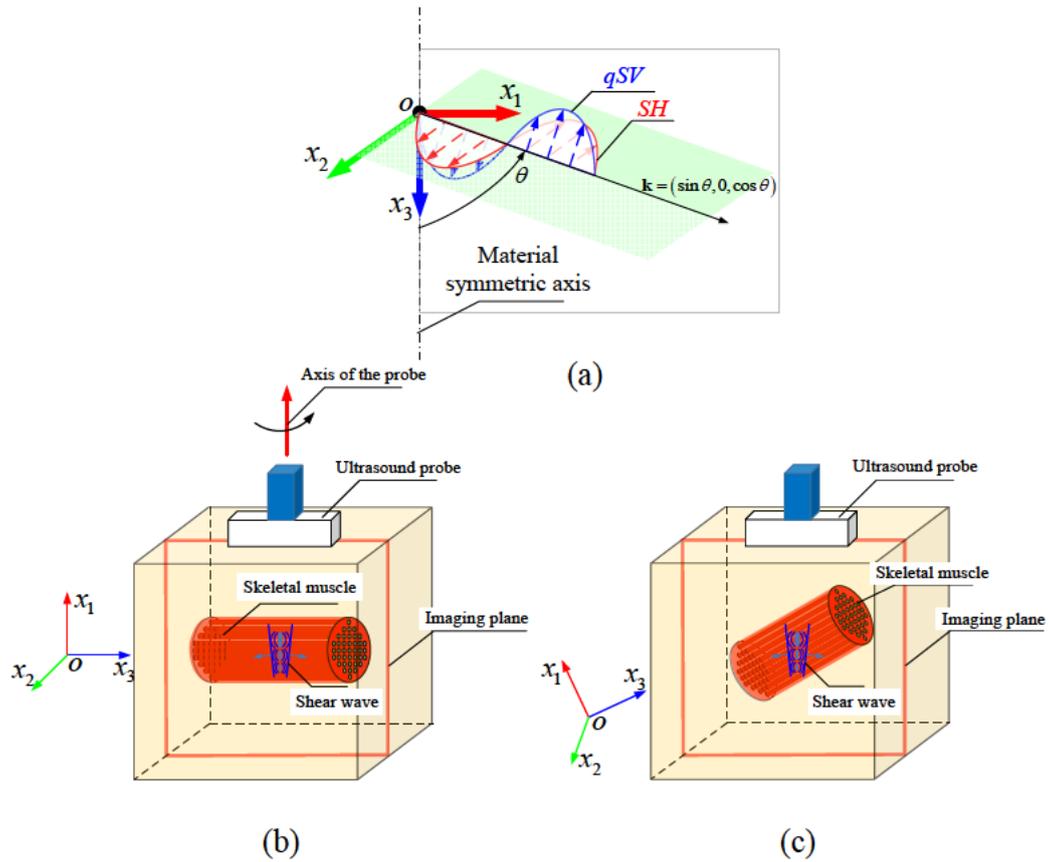

Fig.2 .The schematic of shear waves of the SH and qSV mode and the measurements of skeletal muscle using SSI technique. (a) the polarization direction of the SH wave has no components in $x_3$-axis while the polarization direction of the qSV wave does not. $\theta$ denotes the angle between the wave vector $\mathbf{k}$ and $x_3$-axis, thus $\mathbf{k}=(\sin\theta,0,\cos\theta)$. (b) Using the shear waves of SH mode to measure the elastic parameters of the skeletal muscle. By rotating the ultrasound probe, the velocities of the SH mode shear wave in different directions can be measured. (c) The setup of the experiments proposed in this study. The polarization directions of the shear waves are no longer perpendicular to the material symmetric direction and defined as the qSV mode.

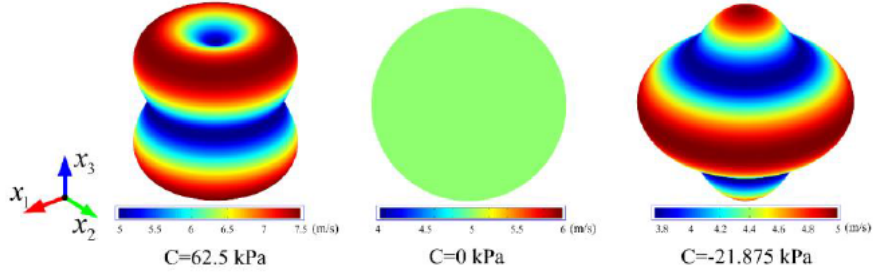

Fig 3. The spatial distribution of the phase velocities for the shear wave of qSV mode. For illustration we take $\mu_T = 9$ kPa, $\mu_L = 25$ kPa for all the three cases, and $C(\text{kPa}) = 62.5, 0, -21.875$, respectively.

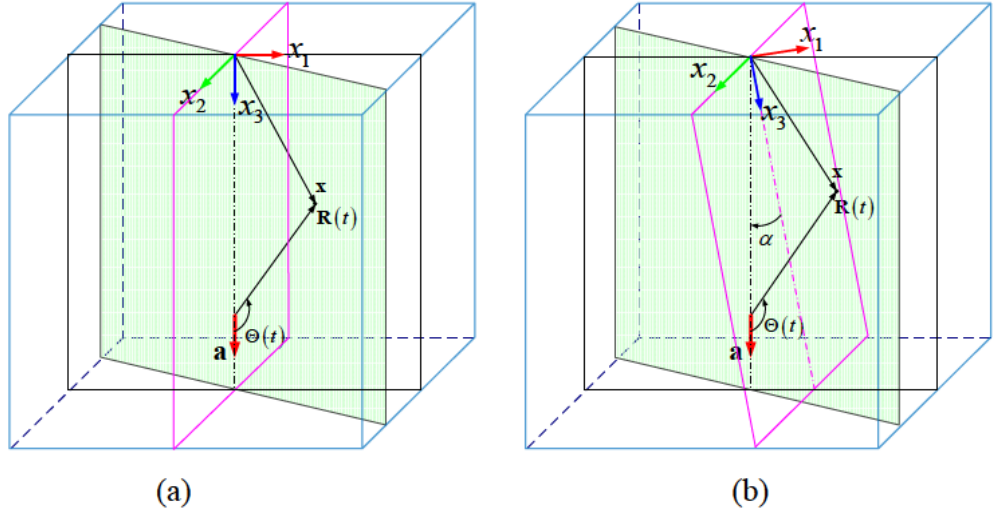

Fig. 4 A schematic of the ECE in a soft media. (a) The ECE in an isotropic elastic solid. The force is moving along $x_3$ axis. The resulting displacements is confined in a Mach cone according to Eq. (3.5). (b) The ECE in an incompressible TI solid. The force is moving along $\mathbf{a} = (-\sin\alpha, 0, \cos\alpha)$ with velocity $v_e$. Here only the Mach cone for the SH mode is plotted. It is no longer a perfect cone for an anisotropic solid in the physical coordinate system.

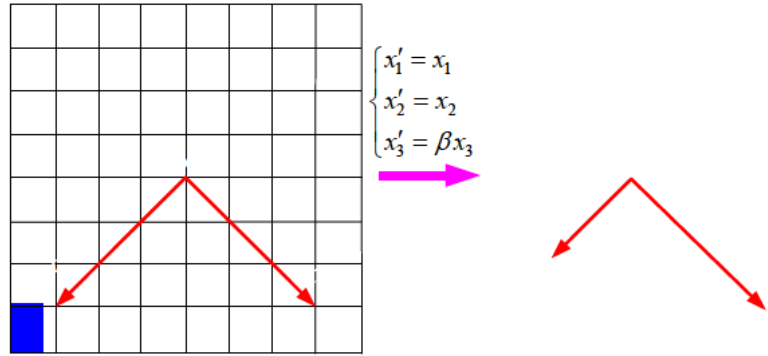

Fig. 5 In the transformed coordinate, the displacement is confined in a Mach cone, $\alpha = 45°$.

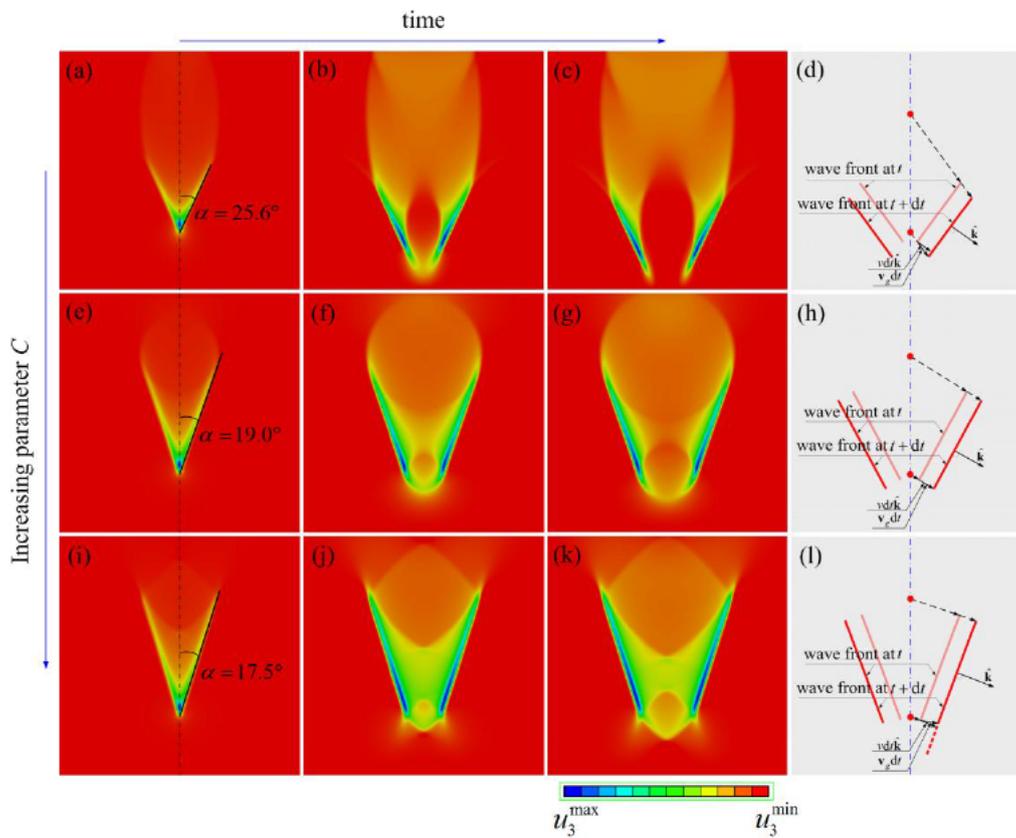

Fig. 6 The ECE at low Mach number when the force is moving along the material symmetric axis direction. The angles of the Mach cones are: (a) $\alpha = 25.6°$, (e) $\alpha = 19.0°$, (i) $\alpha = 17.5°$. (a)-(c) for $C > 0$, (e)-(g) for $C = 0$ and (i)-(k) for $C < 0$, In general, the direction of the wave vector are not in the same directions expect for $C = 0$. Phase velocities are the projections of the group velocities in the directions of the wave vectors as schematically given by (d), (h) and (l).

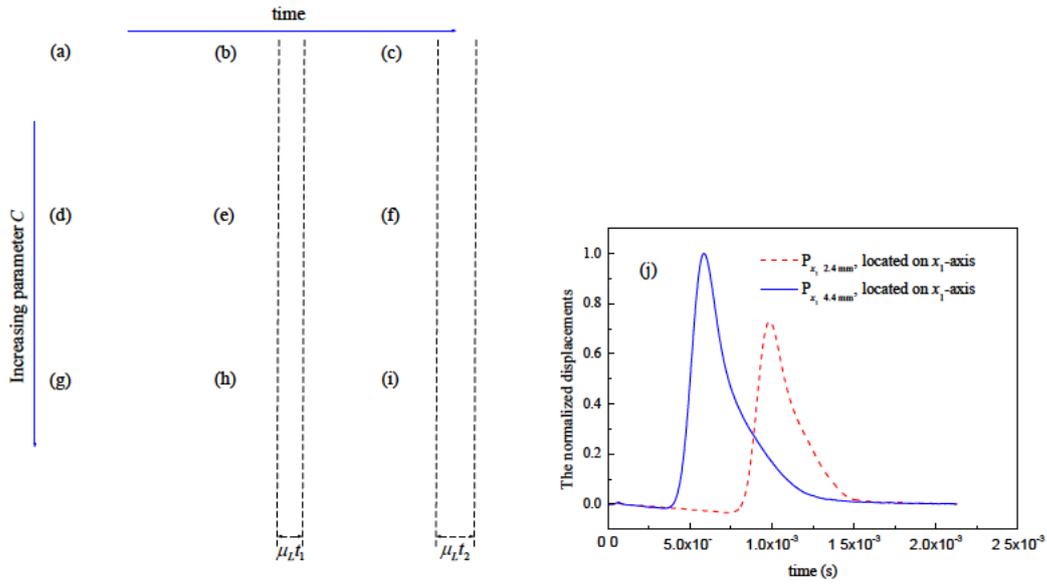

Fig. 7 The ECE at high Mach number when the force is moving along the material symmetric axis direction. The angle of the Mach cone approximately equals to zero and the quasi-plane waves are formed.

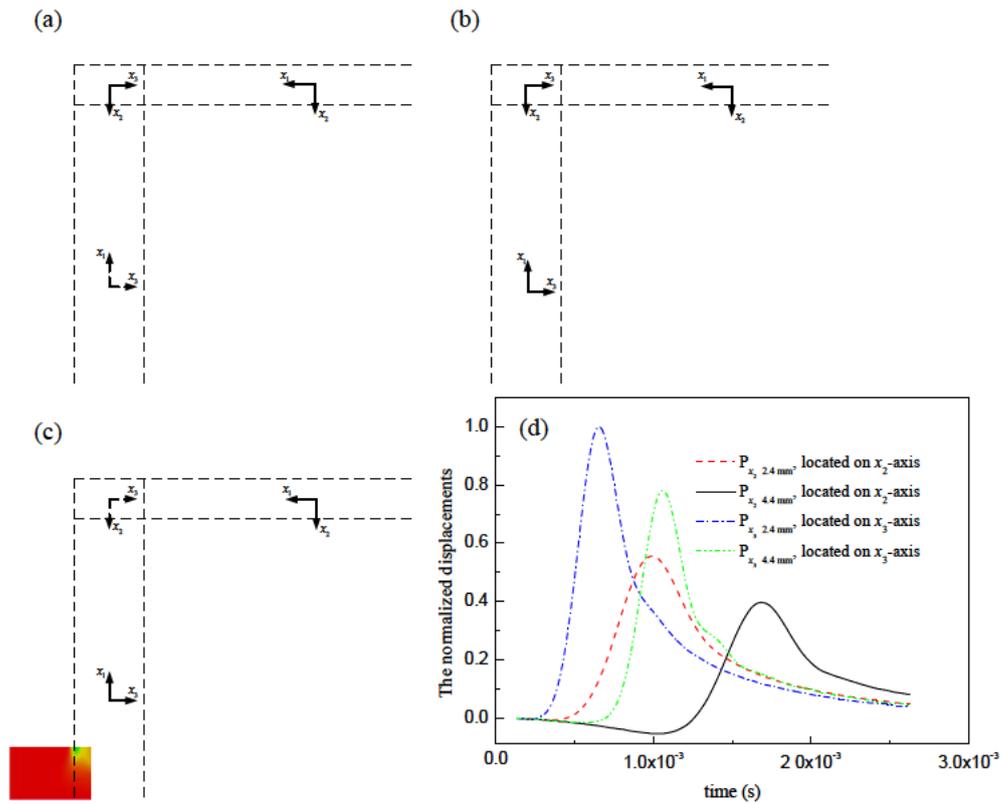

Fig. 8 The ECE at high Mach number when the shear source is moving perpendicular to the material symmetric axis.

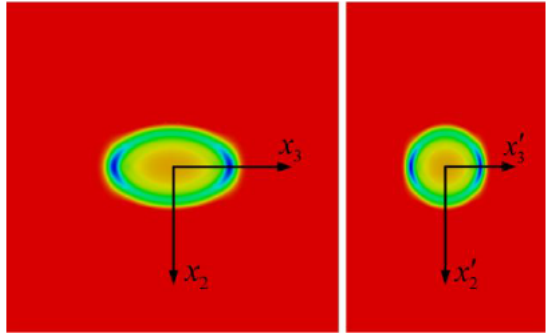

Fig. 9 The Mach cones in the original and the transformed coordinate systems.

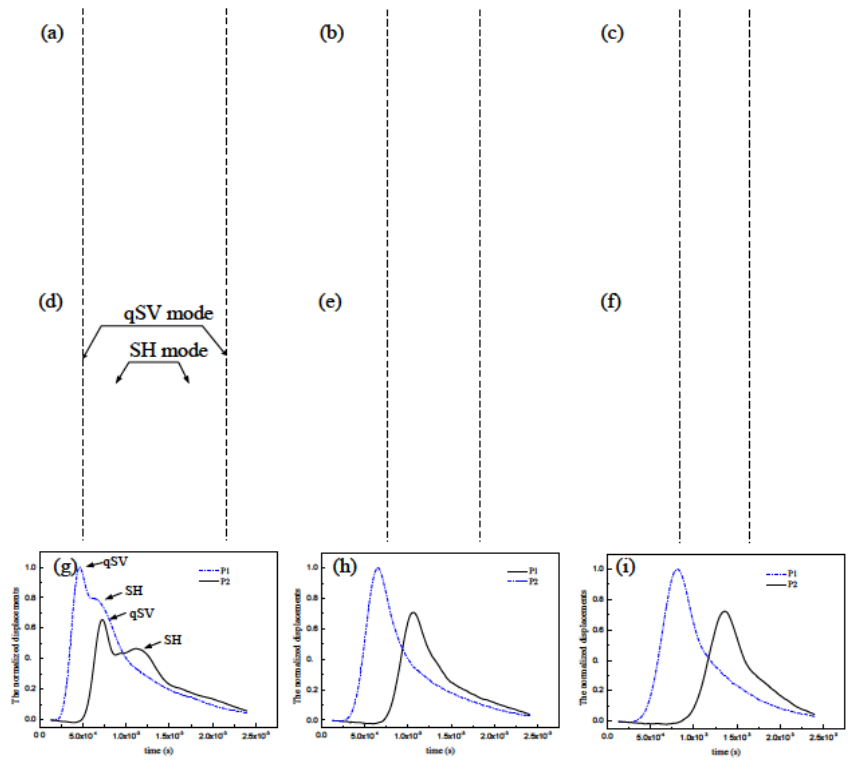

Fig. 10 The ECE at a high Mach number when the angel between the direction of the moving force and the material symmetric axis is taken as $\alpha = 45°$.

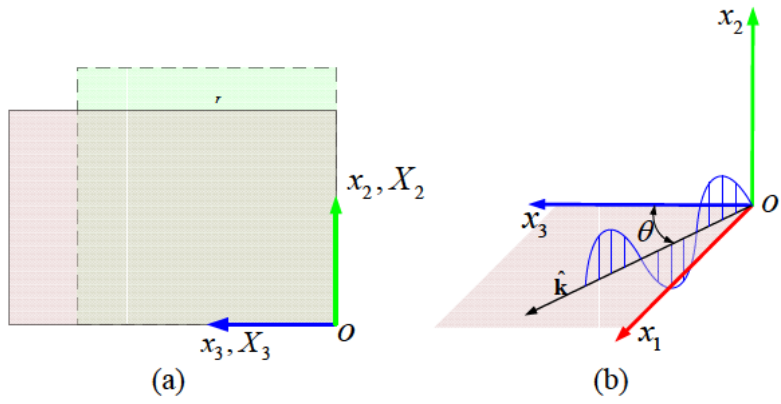

Fig. 11 (a) Illustrate of the original and deformed configurations of the region of interest; (b) illustration of the wave vector

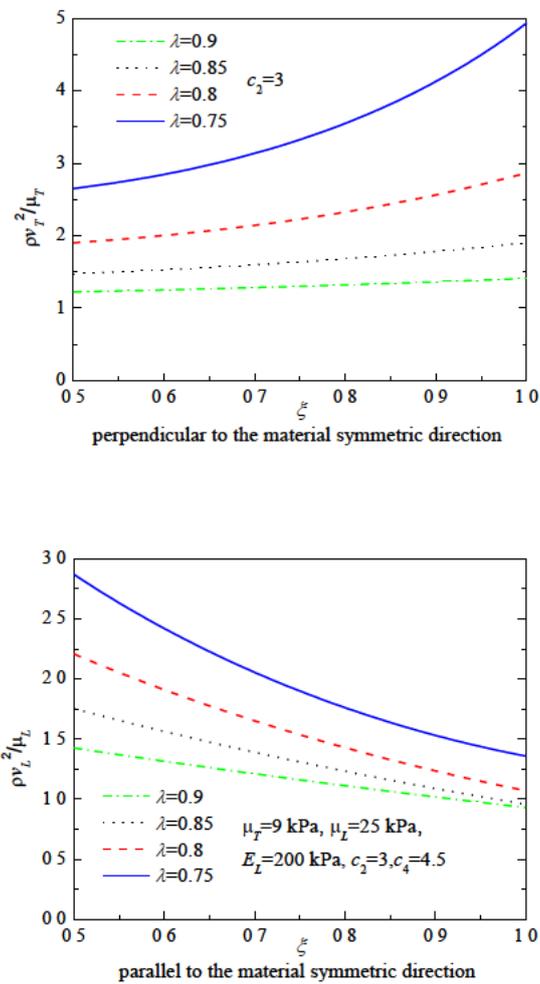

Fig. 12 Dependence of the shear wave velocities $v_L$ and $v_T$ on the parameter $\xi$ for different $\lambda$

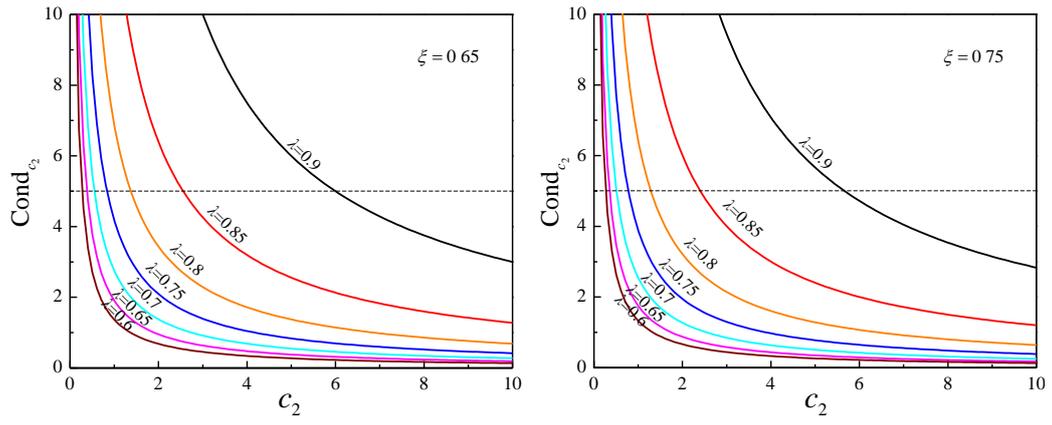

Fig. 13　Plot of the condition number for $c_2$

# Supporting information

## SI-1: Correlations between the elastic tensors and $\mu_L$, $\mu_T$ and $E_L$

The relationships between the linear elastic stiffness or compliance tensor components and the three elastic parameters $\mu_L$, $\mu_T$ and $E_L$ for incompressible transversely isotropic elastic materials are established here.

The constitutive equation for a linear elastic anisotropic solid is, equivalently,

$$\boldsymbol{\sigma} = \boldsymbol{C}\boldsymbol{\varepsilon}, \quad \boldsymbol{\varepsilon} = \boldsymbol{S}\boldsymbol{\sigma} \tag{S1-1}$$

where $\boldsymbol{\sigma} = [\sigma_{11}, \sigma_{22}, \sigma_{33}, \sigma_{23}, \sigma_{31}, \sigma_{12}]^T$ is the infinitesimal stress, $\boldsymbol{\varepsilon} = [\varepsilon_{11}, \varepsilon_{22}, \varepsilon_{33}, 2\varepsilon_{23}, 2\varepsilon_{31}, 2\varepsilon_{12}]^T$ is the infinitesimal strain, $\mathbf{C}$ is the stiffness tensor and $\mathbf{S}$ the compliance tensor.

In the most general case, $\mathbf{C}$ and $\mathbf{S}$ have 21 independent components each. For transversely isotropic (TI) materials, the number of independent components reduces to five due to the existence of material symmetries (Ting, 1996), and $\mathbf{C}$ reads (using Voigt contracted notation)

$$\mathbf{C} = \begin{bmatrix} C_{11} & C_{11}-2C_{66} & C_{13} & 0 & 0 & 0 \\ & C_{11} & C_{13} & 0 & 0 & 0 \\ & & C_{33} & 0 & 0 & 0 \\ & & & C_{44} & 0 & 0 \\ & & & & C_{44} & 0 \\ & & & & & C_{66} \end{bmatrix}. \tag{S1-2}$$

Conversely, the compliance matrix $\mathbf{S}$ can be written as

$$S = \begin{bmatrix} S_{11} & S_{11} - \frac{1}{2}S_{66} & S_{13} & 0 & 0 & 0 \\ & S_{11} & S_{13} & 0 & 0 & 0 \\ & & S_{33} & 0 & 0 & 0 \\ & & & S_{44} & 0 & 0 \\ & & & & S_{44} & 0 \\ & & & & & S_{66} \end{bmatrix} \qquad (S1\text{-}3)$$

with the connections

$$S_{11} = \frac{C_{11}C_{33} - C_{13}^2}{4C_{66}(C_{11}C_{33} - C_{13}^2 - C_{33}C_{66})}, \quad S_{13} = \frac{-C_{13}}{2(C_{11}C_{33} - C_{13}^2 - C_{33}C_{66})},$$

$$S_{33} = \frac{C_{11} - C_{66}}{C_{11}C_{33} - C_{13}^2 - C_{33}C_{66}}, \quad S_{44} = \frac{1}{C_{44}}, \quad S_{66} = \frac{1}{C_{66}}$$

Incompressibility requires that

$$\varepsilon_{11} + \varepsilon_{22} + \varepsilon_{33} = 0, \qquad (S1\text{-}4)$$

and from Eq.(S1-1) leads to the following two relationships between the compliances (Destrade et al., 2002),

$$2S_{11} - \frac{1}{2}S_{66} + S_{13} = 0, \quad 2S_{13} + S_{33} = 0.$$

A linear incompressible TI solid possesses thus only three independent material parameters. Note that the compliance matrix is now positive semi-definite, and no longer invertible. The strain-stress relation in (S1-1) remains unchanged.

In terms of stiffnesses we now have the corresponding limits

$$\frac{C_{13} - C_{33}}{C_{11}C_{33} - C_{13}^2 - C_{33}C_{66}} \to 0, \qquad \frac{C_{11} - C_{13} - C_{66}}{C_{11}C_{33} - C_{13}^2 - C_{33}C_{66}} \to 0,$$

and the stress-strain relation in (S1-1) is modified to

$$\boldsymbol{\sigma} = \boldsymbol{C}\boldsymbol{\varepsilon} + p\boldsymbol{I},$$

where $p$ is a Lagrange multiplier and $\boldsymbol{I}$ the identity tensor. Note that these limits are consistent with the isotropic case, where $C_{11}, C_{13}$ tend to infinity while $C_{33}, C_{66}$ remain finite.

An alternative notation, favored by engineers, can also be adopted for the compliance tensor of a compressible TI solid, as

$$\mathbf{S} = \mathbf{C}^{-1} = \begin{bmatrix} 1/E_T & -\nu_{TT}/E_T & -\nu_{LT}/E_L & 0 & 0 & 0 \\ & 1/E_T & -\nu_{LT}/E_L & 0 & 0 & 0 \\ & & 1/E_L & 0 & 0 & 0 \\ & & & 1/\mu_L & 0 & 0 \\ & & & & 1/\mu_L & 0 \\ & & & & & 1/\mu_T \end{bmatrix}, \quad \text{(S1-5)}$$

where $\mu_T = \dfrac{E_T}{2(1+\nu_{TT})}$. Here the five independent material parameters can be chosen from the Poisson ratios $\nu_{TT}, \nu_{LT}$, the extension moduli $E_T, E_L$, and the shear moduli $\mu_T, \mu_L$. Their identification with the compliances $S_{ij}$ is clear by comparison of the two expressions. In particular

$$E_L = \frac{1}{S_{33}}, \quad \mu_L = \frac{1}{S_{44}} = C_{44}, \quad \mu_T = \frac{1}{S_{66}} = C_{66}.$$

The incompressibility compliance relations now read

$$\begin{cases} 1 - \nu_{TT} - \dfrac{E_T}{2E_L} = 0 \\ \nu_{LT} = \dfrac{1}{2} \end{cases}. \quad \text{(S1-6)}$$

Without loss of generality, we now may choose $\mu_L$, $\mu_T$, $E_L$ as the three independent parameters. With that choice it is easy to check that the compliance tensor of linear incompressible TI solids reads

$$\mathbf{S} = \begin{bmatrix} \frac{1}{4}\left(\frac{1}{E_L}+\frac{1}{\mu_T}\right) & \frac{1}{4}\left(\frac{1}{E_L}-\frac{1}{\mu_T}\right) & -\frac{1}{2E_L} & 0 & 0 & 0 \\ & \frac{1}{4}\left(\frac{1}{E_L}+\frac{1}{\mu_T}\right) & -\frac{1}{2E_L} & 0 & 0 & 0 \\ & & \frac{1}{E_L} & 0 & 0 & 0 \\ & & & \frac{1}{\mu_L} & 0 & 0 \\ & & & & \frac{1}{\mu_L} & 0 \\ & & & & & \frac{1}{\mu_T} \end{bmatrix} \quad \text{(S1-7)}$$

Chadwick (1989), Papazoglou et al. (2006) and Rouze et al. (2013), established that two transverse waves may propagate in this medium, with speeds given by

$$\begin{cases} \rho v_{SH}^2 = \mu_T\left(\hat{k}_1^2+\hat{k}_2^2\right)+\mu_L \hat{k}_3^2 \\ \rho v_{qSV}^2 = \mu_L + 4\left(\frac{E_L}{E_T}\mu_T - \mu_L\right)\left(1-\hat{k}_3^2\right)\hat{k}_3^2 \end{cases}.$$

However, they did not notice that

$$4\left(\frac{E_L}{E_T}\mu_T - \mu_L\right) = E_L + \mu_T - 4\mu_L,$$

which follows from (S1-7). In the paper we call this quantity $2C$.

Vavryčuk (2001) studied a special kind of compressible TI materials, with stiffness matrix $\mathbf{C}$ of the following form,

$$\mathbf{C} = \begin{bmatrix} C_{11} & C_{11}-2C_{66} & C_{11}-2C_{44} & 0 & 0 & 0 \\ & C_{11} & C_{11}-2C_{44} & 0 & 0 & 0 \\ & & C_{11} & 0 & 0 & 0 \\ & & & C_{44} & 0 & 0 \\ & & & & C_{44} & 0 \\ & & & & & C_{66} \end{bmatrix}. \quad \text{(S1-8)}$$

For this case, the equations of incompressibility read

$$\frac{C_{44}}{4C_{11}C_{44}-4C_{44}^2-C_{11}C_{66}} \to 0, \qquad \frac{2C_{44}-C_{66}}{4C_{11}C_{44}-4C_{44}^2-C_{11}C_{66}} \to 0,$$

and because $C_{44} = \mu_L$ and $C_{66} = \mu_T$ must remain finite and >0, we conclude that $C_{11} \to \infty$ in the incompressible limit, consistent with the isotropic case. By inverting $C$, we find that

$$E_L + \mu_T - 4\mu_L = -\frac{(2C_{44} - C_{66})^2}{C_{11} - C_{66}} \quad \text{(S1-9)}$$

indicating that the parameter $C$, defined as

$$C = \frac{E_L + \mu_T - 4\mu_L}{2}. \quad \text{(S1-10)}$$

is zero in the incompressible limit. Therefore this special linear incompressible TI material has only two independent material parameters.

For this type of special solids, Vavryčuk (2001) obtained the *Green function* in closed form. Introducing the density weighted stiffnesses $a_{ij} = C_{ij}/\rho$ $(i, j = 1, 2, 3)$, he found the following explicit expression of the *Green function* $G_{im}^{TI}(\mathbf{x}, t)$

$$G_{im}^{TI}(\mathbf{x}, t) = \frac{1}{4\pi\rho} \left\{ \frac{1}{\sqrt{a_{11}^3}} \frac{g_{1i} g_{1m}}{\tau_1} \delta(t - \tau_1) + \frac{1}{\sqrt{a_{44}^3}} \frac{g_{2i} g_{2m}}{\tau_2} \delta(t - \tau_2) \right.$$
$$+ \frac{1}{a_{66}\sqrt{a_{44}}} \frac{g_{3i} g_{3m}}{\tau_3} \delta(t - \tau_3) + \frac{1}{\sqrt{a_{44}}} \frac{g_{3i}^\perp g_{3m}^\perp - g_{3i} g_{3m}}{r_\perp^2} \quad \text{(S1-11)}$$
$$\left. \times \int_{\tau_3}^{\tau_3} \delta(t - \tau) d\tau + \frac{3 g_{1i} g_{1m} - \delta_{im}}{r^3} \int_{\tau_1}^{\tau_2} \tau \delta(t - \tau) d\tau \right\}$$

where $r = |\mathbf{x}|$, $\gamma_i = \frac{x_i}{r} = \frac{\partial r}{\partial x_i} (i = 1, 2, 3)$, $r_\perp = \sqrt{x_1^2 + x_2^2}$, and

$$\mathbf{g}_1 = \begin{bmatrix} \gamma_1 \\ \gamma_2 \\ \gamma_3 \end{bmatrix}, \mathbf{g}_2 = \frac{-1}{\sqrt{\gamma_1^2 + \gamma_2^2}} \begin{bmatrix} -\gamma_1 \gamma_3 \\ -\gamma_2 \gamma_3 \\ \gamma_1^2 + \gamma_2^2 \end{bmatrix}, \mathbf{g}_3 = \frac{1}{\sqrt{\gamma_1^2 + \gamma_2^2}} \begin{bmatrix} \gamma_2 \\ -\gamma_1 \\ 0 \end{bmatrix},$$
$$\mathbf{g}_3^\perp = \frac{1}{\sqrt{\gamma_1^2 + \gamma_2^2}} \begin{bmatrix} \gamma_1 \\ \gamma_2 \\ 0 \end{bmatrix} \quad \text{(S1-12)}$$

$$\tau_1 = \frac{r}{\sqrt{a_{11}}}, \tau_2 = \frac{r}{\sqrt{a_{44}}}, \tau_3 = \frac{r}{\sqrt{a_{66}}} \sqrt{\gamma_1^2 + \gamma_2^2 + \frac{a_{66}}{a_{44}} \gamma_3^2}, \quad \text{(S1-13)}$$

Taking the incompressible limit, and neglecting the coupling components (Aki and Richards, 2002), we find the approximate form of the *Green function* as

$$G_{im}^{TI}(\mathbf{x},t) \approx \frac{1}{4\pi\rho^{-1/2}}\left\{\frac{1}{\sqrt{\mu_L^3}}\frac{g_{2i}g_{2m}}{\tau_2}\delta(t-\tau_2) + \frac{1}{\mu_T\sqrt{\mu_L}}\frac{g_{3i}g_{3m}}{\tau_3}\delta(t-\tau_3)\right\}, \quad \text{(S1-14)}$$

**SI-2: Elastic Cerenkov effects in an isotropic solid**

The displacement induced by a moving point force can be expressed as

$$u_i(\mathbf{x},t) = \iint a_m \delta(\mathbf{x} - v_e t\mathbf{a}) G_{im}(\mathbf{x}-\boldsymbol{\xi}, t-\tau) \mathrm{d}\tau \mathrm{d}\boldsymbol{\xi}, \qquad (\text{S2-1})$$

where $G_{im}(\mathbf{x},t)$ is the *Green function tensor*. Once we obtain the explicit expression for $G_{im}(\mathbf{x},t)$, the displacement field caused by the moving point force can be deduced from Eq.(S2-1).

First consider isotropic media. The *Green function* can be then written as

$$G_{im}^{Iso} \approx \frac{1}{4\pi\rho v_s^2}(\delta_{im} - \gamma_i \gamma_m)\frac{1}{r}\delta\left(t - \frac{r}{v_s}\right). \qquad (\text{S2-2})$$

Inserting Eq.(S2-2) into Eq.(S2-1) gives

$$u_i(\mathbf{x},t) = \iint \frac{a_m \delta(\boldsymbol{\xi} - v_e \tau \mathbf{a})}{4\pi\rho v_s^2 |\mathbf{x}-\boldsymbol{\xi}|}\left(\delta_{im} - \frac{\partial |\mathbf{x}-\boldsymbol{\xi}|}{\partial x_i}\frac{\partial |\mathbf{x}-\boldsymbol{\xi}|}{\partial x_m}\right)\delta\left(t - \tau - \frac{|\mathbf{x}-\boldsymbol{\xi}|}{v_s}\right)\mathrm{d}\tau\mathrm{d}\boldsymbol{\xi}. \qquad (\text{S2-3})$$

To calculate the integral in Eq.(S2-3), we invoke the following properties of the Kronecker $\delta$-function for arbitrary functions $f(t)$ and $g(t)$,

$$\int f(\tau)\delta(t-\tau)\mathrm{d}\tau = f(t), \qquad (\text{S2-4})$$

and

$$\int f(\tau)\delta(g(\tau))\mathrm{d}\tau = \sum_{i=1}^{N}\left[\frac{f}{|\mathrm{d}g/\mathrm{d}\tau|}\right]_{\tau=\tau_i^*}, \qquad (\text{S2-5})$$

where $\tau_i^*$ ($i=1,2,\mathrm{L},N$) are the roots of the equation $g(\tau)=0$. The proofs for those equations are standard and can be found in any textbook about *generalized functions*.

Using Eq.(S2-4), Eq. (S2-3) can be written in the following form

$$u_i(\mathbf{x},t) =$$

$$\int \frac{a_m}{4\pi\rho v_s^2 |\mathbf{x}-v_e\tau\mathbf{a}|} \left( \delta_{im} - \frac{\partial |\mathbf{x}-v_e\tau\mathbf{a}|}{\partial x_i} \frac{\partial |\mathbf{x}-v_e\tau\mathbf{a}|}{\partial x_m} \right) \delta\left(t-\tau-\frac{|\mathbf{x}-v_e\tau\mathbf{a}|}{v_s}\right) d\tau. \quad (S2\text{-}6)$$

Without loss of generality, let $\mathbf{a}=(0,0,1)$ and define $R(t)$ and $\Theta(t)$ as

$$\begin{cases} \mathbf{R}(t) = \mathbf{x} - v_e t \mathbf{a} \\ \Theta(t) = \arccos\left(\frac{\mathbf{R}\cdot\mathbf{a}}{R}\right) \end{cases}, \quad (S2\text{-}7)$$

where $R=|\mathbf{R}|$. In Eq.(S2-6), we define $g_{Iso}(\tau) = t-\tau-\frac{|\mathbf{x}-v_e\tau\mathbf{a}|}{v_s}$, so

$$|dg_{Iso}/d\tau| = \left| -1 + \frac{x_3 - v_e\tau}{R(\tau)} \frac{v_e}{v_s} \right|. \quad (S2\text{-}8)$$

According to the property of the $\delta$-function given by Eq.(S2-5), we obtain from Eq.(S2-6)

$$u_i(\mathbf{x},t) = \sum_{k=1}^{2} \frac{1}{4\pi\rho v_s^2 R(\tau_k^{Iso})} \left( \delta_{i3} - \frac{\partial R(\tau_k^{Iso})}{\partial x_i} \frac{\partial R(\tau_k^{Iso})}{\partial x_3} \right) \frac{1}{\left| -1 + \frac{x_3 - v_e\tau_k^{Iso}}{R(\tau_k^{Iso})} \frac{v_e}{v_s} \right|}, \quad (S2\text{-}9)$$

where $\tau_k^{Iso}$ ($k=1,2$) are the roots of equation $g_{Iso}(\tau)=0$. The explicit expressions of $\tau_k^{Iso}$ are

$$\tau_{1,2}^{Iso} = t + \frac{R(t)}{v_s(M_{Iso}^2 - 1)} \left( M_{Iso}\cos\Theta(t) \pm \sqrt{1 - M_{Iso}^2 \sin^2\Theta(t)} \right), \quad (S2\text{-}10)$$

where $M = v_e/v_s$ is the *Mach number*. Inserting the expression of $\tau_k^{Iso}$ ($k=1,2$) into Eq.(S2-9), then the following equation can be deduced

$$u_i(\mathbf{x},t) = \frac{1}{\sqrt{1-M_{Iso}^2 \sin^2\Theta(t)}} F^{Iso}\left(\mathbf{x},t,\tau_1^{Iso},\tau_2^{Iso}\right), \quad (S2\text{-}11)$$

where

$$F^{Iso}\left(\mathbf{x},t,\tau_1^{Iso},\tau_2^{Iso}\right) = \sum_{k=1}^{2}\frac{1}{4\pi\mu R(t)}\left(\delta_{i3} - \frac{\partial R\left(\tau_k^{Iso}\right)}{\partial x_i}\frac{\partial R\left(\tau_k^{Iso}\right)}{\partial x_3}\right). \quad (S2\text{-}12)$$

## SI-3: Derivations of the ECE in the special incompressible TI material with $C = 0$

As seen in SI-1, the Green function for the special incompressible TI material with $C = 0$ is

$$G_{im}^{TI}(\mathbf{x},t) \approx \frac{1}{4\pi\rho^{-1/2}} \left\{ \frac{1}{\sqrt{\mu_L^3}} \frac{g_{2i}g_{2m}}{\tau_2} \delta(t-\tau_2) + \frac{1}{\mu_T\sqrt{\mu_L}} \frac{g_{3i}g_{3m}}{\tau_3} \delta(t-\tau_3) \right\}. \quad \text{(S3-1)}$$

where

$$\begin{cases} \mathbf{g}_2 = \dfrac{-1}{\sqrt{\gamma_1^2+\gamma_2^2}} \left(-\gamma_2\gamma_3, -\gamma_2\gamma_3, \gamma_1^2+\gamma_2^2\right) \\ \mathbf{g}_3 = \dfrac{1}{\sqrt{\gamma_1^2+\gamma_2^2}} \left(\gamma_2, -\gamma_1, 0\right) \end{cases}, \quad \text{(S3-2)}$$

and

$$\begin{cases} \tau_2 = \dfrac{r}{\sqrt{\mu_L/\rho}} \\ \tau_3 = \dfrac{r}{\sqrt{\mu_T/\rho}} \sqrt{\gamma_1^2+\gamma_2^2+\dfrac{\mu_T}{\mu_L}\gamma_3^2} \end{cases}. \quad \text{(S3-3)}$$

In Eq.(S3-1), the first term describes the far-field wave of the qSV mode, while the second term describes the far-field wave of the SH mode. We use $G_{im}^{qSV}(\mathbf{x},t)$ and $G_{im}^{SH}(\mathbf{x},t)$ to represent those two modes, respectively,

$$\begin{cases} G_{im}^{qSV}(\mathbf{x},t) = \dfrac{1}{4\pi\rho^{-1/2}} \dfrac{1}{\sqrt{\mu_L^3}} \dfrac{g_{2i}g_{2m}}{\tau_2} \delta(t-\tau_2) \\ G_{im}^{SH}(\mathbf{x},t) = \dfrac{1}{4\pi\rho^{-1/2}} \dfrac{1}{\mu_T\sqrt{\mu_L}} \dfrac{g_{3i}g_{3m}}{\tau_3} \delta(t-\tau_3) \end{cases}. \quad \text{(S3-4)}$$

Then the displacement fields of the two modes, $u_i^{qSV}$ and $u_i^{SH}$, can be calculated by inserting Eq.(S3-4) and Eq.(3.2) into Eq.(3.1), respectively.

We first derive the formula for the qSV mode. Inserting Eq.(S3-4) and Eq.(3.2) into Eq.(3.1) we have

$$u_i^{qSV} = \frac{a_m}{4\pi\mu_L^{3/2}\rho^{-1/2}} \iint \delta(\boldsymbol{\xi}-v_e\tau\mathbf{a}) \frac{\dot{g}_{2i}\dot{g}_{2m}}{\dot{\tau}_2} \delta(t-\tau-\dot{\tau}_2) d\xi d\tau. \quad \text{(S3-5)}$$

In Eq.(S3-5), the superscript '˜' means the variable is a function of $(\mathbf{x}-\boldsymbol{\xi})$. Using Eq.(S2-4), we have

$$u_i^{qSV} = \frac{a_m}{4\pi\mu_L^{3/2}\rho^{-1/2}}\int \frac{\bar{g}_{2i}\bar{g}_{2m}}{\bar{\tau}_2}\delta(t-\tau-\bar{\tau}_2)d\tau. \qquad (S3-6)$$

In Eq.(S3-6), the superscript '¯' means the variable is a function of $(\mathbf{x}-v_e\tau\mathbf{a})$. To evaluate the integral in Eq. (S3-6), we invoke Eq.(S2-5), for which we need to calculate the roots of $g_{qSV}(\tau)=t-\tau-\bar{\tau}_2=0$, i.e.,

$$\left(v_e^2-\frac{\mu_L}{\rho}\right)\tau^2 - 2\left(x_1 v_{e1}+x_3 v_{e3}-\frac{\mu_L}{\rho}t\right)\tau + \left(x_1^2+x_2^2+x_3^2-\frac{\mu_L}{\rho}t^2\right)=0. \qquad (S3-7)$$

The Mach number $M_{qSV}=v_e/\sqrt{\mu_L/\rho}$ is larger than 1 in this study, i.e., $v_e > \sqrt{\mu_L/\rho}$. Thus the roots of Eq.(S3-7) are

$$\tau_{1,2}^{qSV} = t + \frac{M_{qSV}R(t)\cos\Theta(t)}{(M_{qSV}^2-1)\sqrt{\mu_L/\rho}} \pm \frac{R(t)\sqrt{1-M_{qSV}^2\sin^2\Theta(t)}}{(M_{qSV}^2-1)\sqrt{\mu_L/\rho}}, \qquad (S3-8)$$

where

$$\begin{cases} \mathbf{R}(t) = \mathbf{x}-v_e t\mathbf{a} \\ \Theta(t) = \arccos\left(\frac{\mathbf{R}\cdot\mathbf{a}}{|\mathbf{R}|}\right) \end{cases}. \qquad (S3-9)$$

From Eqs.(S2-5) and (S3-6), we obtain

$$u_i^{qSV} = \sum_{k=1}^{2} \frac{a_m}{4\pi\mu_L^{3/2}\rho^{-1/2}} \left[\frac{\bar{g}_{2i}\bar{g}_{2m}}{\bar{\tau}_2}\frac{1}{|dg_{qSV}/d\tau|}\right]_{\tau=\tau_k^{qSV}} \qquad (S3-10)$$

Inserting Eq.(S3-8) into Eq.(S3-10) and consider $\frac{dg_{qSV}}{d\tau}=-1+\frac{d\bar{\tau}_2}{d\tau}=-1+\frac{1}{\sqrt{\mu_L/\rho}}\frac{dR(\tau)}{d\tau}$, we arrive at

$$u_i^{qSV} = \sum_{k=1}^{2} \frac{a_m}{4\pi\mu_L} \frac{1}{R(t)\sqrt{1-M_{qSV}^2\sin^2\Theta(t)}}(\bar{g}_{2i}\bar{g}_{2m})\Big|_{\tau=\tau_k^{qSV}}, \qquad (S3-11)$$

or

$$u_i^{qSV} = \frac{1}{\sqrt{1-M_{qSV}^2 \sin^2 \Theta(t)}} F^{qSV}\left(\mathbf{x},t,\tau_1^{qSV},\tau_2^{qSV}\right). \tag{S3-12}$$

We now proceed to derive the Eq.(3-18) in the main text. Inserting Eqs.(S3-4) and (3.2) into Eq.(3.1) we have

$$u_i^{SH} = \frac{a_m}{4\pi\mu_T\sqrt{\mu_L}\rho^{-1/2}} \iint \delta(\boldsymbol{\xi}-v_e\tau\mathbf{a})\frac{\overset{\text{\tiny J}}{g}_{3i}\overset{\text{\tiny J}}{g}_{3m}}{\overset{\text{\tiny J}}{t}_3} \delta(t-\tau-\overset{\text{\tiny J}}{t}_3)\mathrm{d}\boldsymbol{\xi}\mathrm{d}\tau. \tag{S3-13}$$

In Eq.(S3-13), the superscript ' $\text{\tiny J}$ ' means the variable is a function of $(\mathbf{x}-\boldsymbol{\xi})$. Invoking Eq.(S2-5), we can obtain

$$u_i^{SH} = \frac{a_m}{4\pi\mu_T\sqrt{\mu_L}\rho^{-1/2}} \int \frac{\bar{g}_{3i}\bar{g}_{3m}}{\bar{\tau}_3} \delta(t-\tau-\bar{\tau}_3)\mathrm{d}\tau. \tag{S3-14}$$

In Eq.(S3-14), the superscript ' $\bar{\phantom{x}}$ ' means the variable is a function of $(\mathbf{x}-v_e\tau\mathbf{a})$. To solve Eq. (S3-14), we need to find the roots of $g_{SH}(\tau) = t-\tau-\bar{\tau}_3 = 0$. To make the form of the solutions more concise, we introduce a coordinate transformation

$$\begin{cases} x_1' = x_1 \\ x_2' = x_2 \\ x_3' = \beta x_3 \end{cases}, \tag{S3-15}$$

where $\beta^2 = \frac{\mu_T}{\mu_L}$. In the transformed coordinates the position vector is defined as $\mathbf{x}' = x_i'\mathbf{e}_i$ and $\gamma_i'$ is defined as $\gamma_i' = \frac{x_i'}{|\mathbf{x}'|}$. The moving direction and speed of the moving source become

$$\mathbf{a}' = \frac{(\sin\alpha, 0, \beta\cos\alpha)}{\sqrt{\sin^2\alpha + \beta^2\cos^2\alpha}}, \tag{S3-16}$$

and

$$v_e' = v_e\sqrt{\sin^2\alpha + \beta^2\cos^2\alpha}, \tag{S3-17}$$

in the transformed coordinates, respectively.

The equation $g_{SH}(\tau) = t - \tau - \bar{\tau}_3 = 0$ now is given by

$$\left(v_e'^2 - \frac{\mu_T}{\rho}\right)\tau^2 - 2\left(x_1'v_{e1}' + x_3'v_{e3}' - \frac{\mu_T}{\rho}t\right)\tau + \left(x_1'^2 + x_2'^2 + x_3'^2 - \frac{\mu_T}{\rho}t^2\right) = 0. \quad \text{(S3-18)}$$

Note that Eq.(S3-18) has basically the same form as Eq.(S3-7). So in the transformed coordinates, we define $\mathbf{R}'(t)$ and $\Theta'(t)$ by referring to Eq.(S3-9), as

$$\begin{cases} \mathbf{R}'(t) = \mathbf{x}' - v_e' t \mathbf{a}' \\ \Theta'(t) = \arccos\left(\frac{\mathbf{R}' \cdot \mathbf{a}'}{|\mathbf{R}'|}\right) \end{cases}. \quad \text{(S3-19)}$$

Then the roots of the Eq.(S3-18) can be expressed as

$$\tau_{1,2}^{SH} = t + \frac{M_{SH} R'(t)\cos\Theta'(t)}{(M_{SH}^2 - 1)\sqrt{\mu_T/\rho}} \pm \frac{R'(t)\sqrt{1 - M_{SH}^2 \sin^2\Theta'(t)}}{(M_{SH}^2 - 1)\sqrt{\mu_T/\rho}}, \quad \text{(S3-20)}$$

where the Mach number is $M_{SH} = v_e'/\sqrt{\mu_T/\rho}$. Here we focus on the supersonic case $M_{SH} > 1$. By invoking the property of $\delta$-function given by Eq.(S2-5), and noticing that $\frac{dg_{SH}}{d\tau} = -1 + \frac{d\bar{\tau}_3}{d\tau} = -1 + \frac{1}{\sqrt{\mu_T/\rho}}\frac{dR'(\tau)}{d\tau}$, we have

$$u_i^{SH} = \sum_{k=1}^{2} \frac{a_m}{4\pi\sqrt{\mu_T \mu_L}} \frac{1}{R'(t)\sqrt{1 - M_{SH}^2 \sin^2\Theta'(t)}} (\bar{g}_{3i}\bar{g}_{3m})\Big|_{\tau = \tau_k^{SH}}. \quad \text{(S3-21)}$$

or

$$u_i^{SH} = \frac{1}{\sqrt{1 - M_{SH}^2 \sin^2\Theta'(t)}} F^{SH}\left(\mathbf{x}', t, \tau_1^{SH}, \tau_2^{SH}\right). \quad \text{(S3-22)}$$